\documentclass[%
reprint,
superscriptaddress,
amsmath,amssymb,
aps,
prb,
prstab,
floatfix,
longbibliography
]{revtex4-2}

\usepackage[caption=false]{subfig}
\usepackage{tabularx}
\usepackage{multirow}
\usepackage{graphicx}% Include figure files
\usepackage{dcolumn}% Align table columns on decimal point
\usepackage{bm}% bold math
\usepackage{threeparttable}
\usepackage{booktabs}
\usepackage{color}
\usepackage{booktabs}
\usepackage{xspace}
\usepackage{chemformula}
\usepackage{xcolor}
\usepackage{scrextend}
\usepackage{hyperref}
\usepackage{orcidlink}
\hypersetup{colorlinks=true, linkcolor=blue, citecolor=blue, urlcolor=blue, pdftitle = {Incipient geometric lattice instability of cubic fluoroperovskites}, pdfauthor = {R.~M.~Dubrovin}}

\definecolor{myblue}{RGB}{0, 0, 0}
\definecolor{newtext}{RGB}{0, 0, 0}

\bibpunct{[}{]}{;}{n}{}{}

\begin{document}

% \title{Incipient geometric lattice instability of cubic inorganic metal halide perovskites}
\title{Incipient geometric lattice instability of cubic \textcolor{newtext}{fluoroperovskites}}

\author{R.~M.~Dubrovin\,\orcidlink{0000-0002-7235-7805}}
\email{dubrovin@mail.ioffe.ru}
\affiliation{Ioffe Institute, Russian Academy of Sciences, 194021 St.\,Petersburg, Russia}
\author{A.~C.~Garcia-Castro\,\orcidlink{0000-0003-3379-4495}}
\affiliation{School of Physics, Universidad Industrial de Santander, Carrera 27 Calle 9, 680002 Bucaramanga, Colombia}
\author{N.~V.~Siverin\,\orcidlink{0000-0002-4643-845X}}
\affiliation{Ioffe Institute, Russian Academy of Sciences, 194021 St.\,Petersburg, Russia}
\author{N.~N.~Novikova\,\orcidlink{0000-0003-2428-6114}}
\affiliation{Institute of Spectroscopy, Russian Academy of Sciences, 108840 Moscow, Troitsk, Russia}
\author{K.~N.~Boldyrev\,\orcidlink{0000-0002-3784-7294}}
\affiliation{Institute of Spectroscopy, Russian Academy of Sciences, 108840 Moscow, Troitsk, Russia}
\author{Aldo~H.~Romero\,\orcidlink{0000-0001-5968-0571}}
\affiliation{Department of Physics and Astronomy, West Virginia University, WV-26506-6315 Morgantown, West Virginia, USA}
\author{R.~V.~Pisarev\,\orcidlink{0000-0002-2008-9335}}
\affiliation{Ioffe Institute, Russian Academy of Sciences, 194021 St.\,Petersburg, Russia}

\date{\today}

\begin{abstract}
Inorganic metal halide perovskites are promising materials for next-generation technologies due to plethora of unique physical properties, many of which cannot be observed in the oxide perovskites.
On the other hand, the search for ferroelectricity and multiferroicity in lead-free inorganic halide perovskites remains a challenging research topic.
Here, we experimentally show that cubic fluoroperovskites exhibit proximity to incipient ferroelectrics, which manifested in the softening of the low-frequency polar phonons in the Brillouin zone center at cooling.
% Here, we experimentally show that the cubic fluoroperovskites exhibit proximity to incipient ferroelectric instability manifested in the softening of the low-frequency polar phonons in the Brillouin zone center at cooling. % in a decrease of the polar phonons frequencies in the Brillouin zone center at cooling.
Furthermore, we reveal the coupling between harmonic and anharmonic force constants of the softening phonons and their correlation with the perovskite tolerance factor.
Next, using first-principles calculations, we examine the lattice dynamics of the cubic fluoroperovskites and disclose the incipient lattice instability at which the harmonic force constants of low-lying phonons tend to decrease with a reduction of tolerance factor at all high-symmetry points of the Brillouin zone.
The correlations with the tolerance factor indicate the geometric origin of observed incipient lattice instability in the cubic fluoroperovskites caused by the steric effect due to the volume filling of the unit cell by different radius ions.
These results provide insights into the lattice dynamics and potential ferroelectric properties of inorganic lead-free metal halide perovskites, relevant to further design and synthesis of new multifunctional materials.
\end{abstract}

\maketitle

\section{Introduction}
Perovskites, primarily oxides, have been having a tremendous impact on physics and technology for many years due to a plethora of intriguing physical phenomena such as magnetism, ferroelectricity, multiferroicity, superconductivity, piezoelectricity, colossal magnetoresistance, etc~\cite{spaldin2019advances}.
Some of these phenomena arise from the flexible crystal structure and the wide variety of elements that perovskites may contain.
In the past decade, metal halide perovskites \ch{AMX3} with a monovalent inorganic or organic \ch{A^{1+}} cation, a divalent metal \ch{B^{2+}} cation, and a halide \ch{X^{1-}} anion such as \ch{F^{1-}}, \ch{Cl^{1-}}, \ch{Br^{1-}}, or \ch{I^{1-}} have become an appealing class of materials for next-generation high-performance optoelectronic, photonic, spintronic devices and beyond due to potentially unique optical and electrical properties that can not be observed in oxide perovskites~\cite{ricciardulli2021emerging,neumann2021manganese,cherniukh2021perovskite,long2020chiral,fu2019metal,snaith2018present}.
On the other hand, many physical phenomena observed in oxide perovskites can also be present in metal halide perovskites.

Ferroelectric properties were discovered in some organic metal halide perovskites, e.g. in \ch{CH3NH3PbI3}~\cite{shahrokhi2020emergence}, and in inorganic ones, such as orthorhombic \ch{CsPbBr3}~\cite{li2020evidence}, and hexagonal \ch{KNiCl3}~\cite{machida1994ferroelectricity}, \ch{TlFeCl3}~\cite{yamanaka2002structural}, \ch{RbMnBr3}~\cite{kato1994successive}, \ch{RbCoBr3}~\cite{morishita2000dielectric}, and \ch{RbFeBr3}~\cite{mitsui1994ferroelectric}.
However, none of the synthesized inorganic metal halide fluoroperovksites \ch{AMF3} was reported to be ferroelectric with the only exception being \ch{CsPbF3} with a lone pair active \ch{Pb^{2+}} cation which was experimentally observed in a non-centrosymmetric space group $R3c$~\cite{berastegui2001low,smith2015interplay}.
Moreover, although the polar ground state was theoretically predicted in \ch{NaCaF3}, \ch{NaCdF3}, \ch{LiMgF3}, and \ch{LiNiF3} fluoroperovskites~\cite{edwardson1989ferroelectricity,claeyssens2003abinitio,duan2004electronic}, up to date, there have been no reports on the synthesis of these crystals. %no reports on the synthesis of these crystals have been observed.

Notwithstanding the foregoing, it was theoretically predicted in Ref.~\cite{garcia2014geometric} that synthesized orthorhombic fluoroperovskites have ferroelectric instability in their hypothetical high-symmetry cubic phase with the degree of which correlates with the tolerance factor $t$~\footnote{The Goldschmidt tolerance factor $t$ is a dimensionless parameter defined as $t = \cfrac{r_{\ch{A}} + r_{\ch{F}}}{\sqrt{2} (r_{\ch{B}} + r_{\ch{F}})}$, where $r_{\ch{A}}$, $r_{\ch{B}}$ and $r_{\ch{F}}$ are ionic radii and which is used for predicting the stability of the fluoroperovskite crystal structure.
Fluoroperovskites with \mbox{$t<0.78$} have the trigonal structure $R3c$ (\#161, $Z=6$), in the range of \mbox{$0.78<t<0.88$} the orthorhombic phase $Pnma$ (\#62, $Z=4$) is stabilized, the cubic phase $Pm\overline{3}m$ (\#221, $Z=1$) is adopted in the range of \mbox{$0.88<t<1.00$}, and the hexagonal structure $P6_{3}/mmc$ (\#194, $Z=6$) is realized for \mbox{$1.00<t<1.08$}~\cite{babel1967structural}.
There are a few exceptions, e.g., \ch{KCuF3}\xspace and \ch{KCrF3}\xspace in which the tetragonal $I4/mcm$ structure is realized due to the Jahn-Teller distortions.
}\nocite{babel1967structural}.
It is worth noting that the ferroelectric instability in fluoroperovskites originates from geometric ionic size effects without noticeable hybridization, which plays a crucial role in the oxide perovskites~\cite{garcia2014geometric}.
The latter prediction was experimentally corroborated and it has been observed that the bulk crystal of the orthorhombic fluoroperovskite \ch{NaMnF3}, with the lowest tolerance factor $t=0.78$, is an incipient multiferroic in which incipient ferroelectricity coexists and even interacts with antiferromagnetic ordering below the N{\'e}el temperature $T_{N}=66$\,K~\cite{dubrovin2020incipient}.
Furthermore, it was shown that the strained thin film of \ch{NaMnF3} is ferroelectric already at room temperature~\cite{garcia2016strain,yang2017room}. 
This intriguing results put on the agenda the necessity of further detailed experimental studies of the lattice dynamics of cubic fluoroperovskites with different tolerance factors $t$ aiming to unveil any signs of incipient ferroelectricity in the high-symmetry cubic phase.

In this paper, we report results of systematic study of the lattice dynamic of cubic fluoroperovskites by far-infrared spectroscopy technique supported by appropriate theoretical calculations and analysis. 
We experimentally revealed that the low-frequency polar phonon softens with cooling at the $\Gamma$ point of the Brillouin zone in all studied crystals, similar to what is observed in incipient ferroelectrics. 
This frequency change correlates with the tolerance factors $t$ of cubic fluoroperovskites so that the lower $t$, the more significant frequency decrease at cooling is observed.
The coupling between harmonic and anharmonic force constants of polar softening phonons is experimentally observed in these crystals.
Moreover, according to our harmonic first-principles simulations, the cubic fluoroperovskites tend to lattice softening at all high-symmetry points of the Brillouin zone with a reduction of the tolerance factor $t$ that indicates to geometric origin of this effect.
% We usually do not add Section descriptions to papers less than 20 pages.

\section{Methods}
\begin{figure}
\centering
\includegraphics[width=\columnwidth]{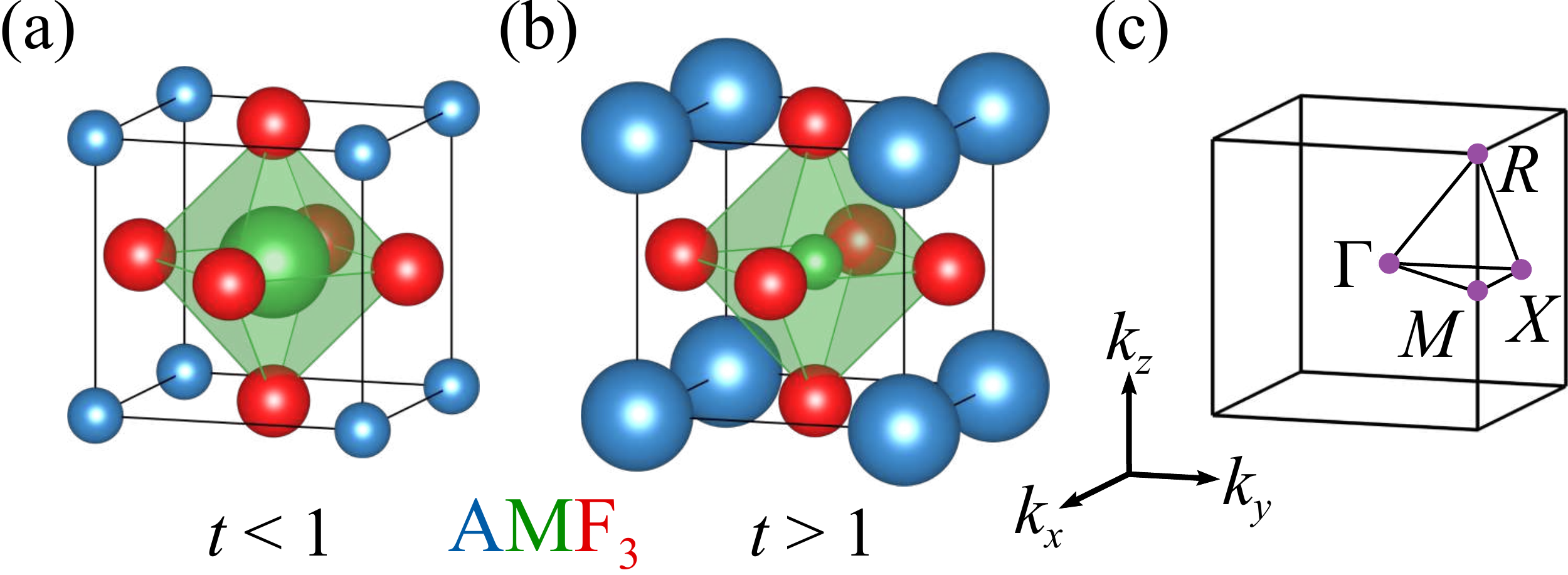}
\caption{\label{fig:structure}
The crystal structure of cubic fluoroperovskites \ch{AMF3} with the tolerance factor (a)~$t<1$ and (b)~$t>1$.
(c)~The Brillouin zone of a face-centered cubic lattice indicating the high-symmetry $\Gamma$ -- $M$ -- $R$ -- $\Gamma$ -- $X$ path used in the lattice dynamic calculations.
The $k_{x}$, $k_{y}$, and $k_{z}$ are the primitive reciprocal lattice vectors.
}
\end{figure}

The experimentally studied cubic fluoroperovskites \ch{KZnF3}, \ch{RbMnF3}, \ch{KNiF3}, and \ch{KMgF3} have the crystal structure which belongs to the space group $Pm\overline{3}m$ (\#221, $Z=1$)~\cite{knight2017low,knox1961perovskite,okazaki1961crystal,windsor1966spin,vaitheeswaran2007high}.
The lattice parameters $a$ of these crystals at room temperature are listed in Table~\ref{tab:phonon_parameters}.
The perovskite unit cell contains five ions occupying the Wyckoff positions $1a$ (0, 0, 0) for \ch{A^1+}\xspace, $1b$ ($\frac{1}{2}$, $\frac{1}{2}$, $\frac{1}{2}$) for \ch{M^2+}\xspace, and $3c$ (0, $\frac{1}{2}$, $\frac{1}{2}$) for \ch{F^1-}\xspace as shown in Figs.~\ref{fig:structure}(a) and~~\ref{fig:structure}(b).
The fluoroperovskites \ch{KZnF3} and \ch{KMgF3} are diamagnetic, while \ch{RbMnF3}  and \ch{KNiF3} are antiferromagnetic below N{\'e}el temperatures $T_{N}=83.5$\,K~\cite{lopez2014magnetic} and 244.8\,K~\cite{nouet1972determination,oleaga2015critical}, respectively.

Single crystals of cubic fluoroperovskites were grown by the Czochralsky method~\cite{gesland1980growth}.
The x-ray oriented crystals were cut normal to the $a$ axis and optically polished.
The surface size of the samples used in the far-infrared experiments were about $10\times{10}$\,mm$^{2}$.
Samples for dielectric measurements were prepared in a form of plane-parallel optically polished plates with a thickness about 0.5\,mm and area about 10\,mm$^{2}$.

The far-infrared (IR) reflectivity measurements were carried out in the spectral range of 30--700\,cm$^{-1}$ with near normal incident light (the incident light beam was at 10$^{\circ}$ from the normal to the crystal surface) using Bruker~IFS~125HR spectrometer equipped by a liquid helium cooled bolometer as a detector.
Due to the cubic symmetry of crystals, three axes are equivalent, and reflectivity measurements were performed with unpolarized light.
Samples were attached to a cold finger of a closed cycle helium cryostat Cryomech~ST403 and the relative reflectivity spectra were measured at continuous cooling from 300 to 5\,K with respect to a reference reflectivity of a gold mirror at room temperature.
No corrections on the surface quality and shape of sample, as well as the sample positions due to cold finger thermal contraction were done.
The absolute reflectivity spectra were obtained at room temperature using Bruker~IFS~66v/S spectrometer in the range of 50--7500\,cm$^{-1}$ with DTGS (50--450\,cm$^{-1}$) and DLaTGS (450--7500\,cm$^{-1}$) detectors which allowed us to determine the high-frequency dielectric permittivity $\varepsilon_{\infty}$.
According to Refs.~\cite{markovin1976observation,krichevtsov1984isotropic} the values of $\varepsilon_{\infty}$ in the cubic fluoroperovskites are characterized by weak temperature changes, which were neglected in the analysis. 

Measurements of the low-frequency dielectric permittivity $\varepsilon^{\textrm{lf}}_{0}$ were done in the range from 20\,Hz to 1\,MHz using precision RLC meter AKTAKOM AM-3028.
Electric contacts were deposited on the sample faces using silver paint to form a capacitor.
Samples were placed in a helium flow cryostat Cryo CRC-102 and measurements were performed at continuous heating from 5 to 300\,K.
Experimental data are presented only at 100\,kHz because no noticeable frequency dispersion was observed.
The dielectric losses were very small, on the order of 10$^{-5}$, with no perceptible temperature changes.

\begin{figure*}
\centering
\includegraphics[width=2\columnwidth]{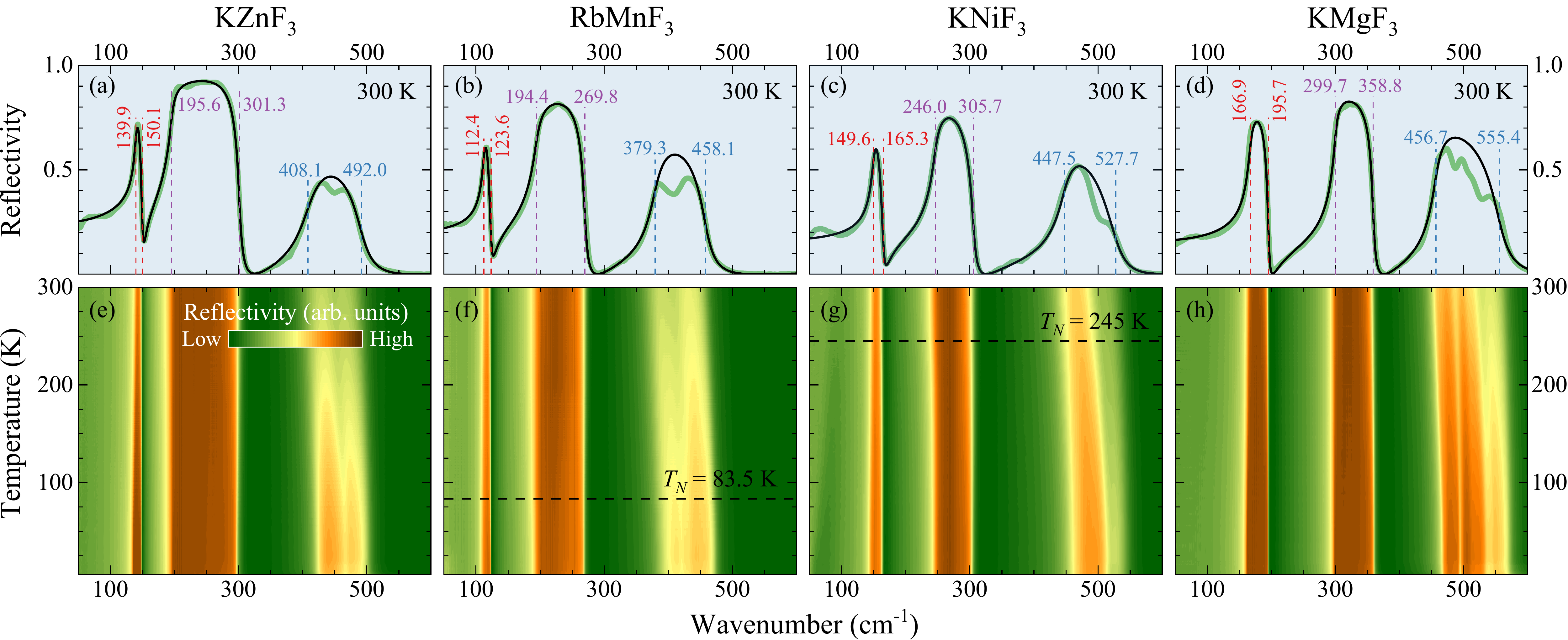}
\caption{\label{fig:reflectivity}
(a)--(d) Room temperature and (e)--(h) temperature colormap of far-infrared reflectivity spectra of the cubic fluoroperovskites \ch{KZnF3}, \ch{RbMnF3}, \ch{KNiF3}, and \ch{KMgF3}, respectively.
The solid black lines are results of fits based on the generalized oscillator model according to Eq.~\eqref{eq:epsilon_TOLO}.
Color vertical dashed lines indicate $\omega_{\textrm{TO}}$ and $\omega_{\textrm{LO}}$ phonon frequencies, with $\omega_{\textrm{TO}}<\omega_{\textrm{LO}}$.
Horizontal black dashed lines indicate antiferromagnetic phase transition temperatures $T_{N}$.
}
\end{figure*}

The experimental results were supported by the use of first-principles calculations of the lattice dynamics within the density functional theory (DFT) framework~\cite{hohenberg1964inhomogeneous,kohn1965self} as implemented in the \textsc{vasp} code~\cite{kresse1996efficient,kresse1999from}.
The projected augmented wave (PAW) method~\cite{blochl1994projector} was used to represent the valence and core electrons.
The following electronic configurations of valence electrons of \ch{K} ($3p^{6}4s^{1}$, version 17Jan2003), \ch{Rb} ($4p^{6}5s^{1}$, version 06Sep2000), \ch{Ca} ($3s^{2}3p^{6}4s^{2}$, version 06Sep2000), \ch{Mn} ($3p^{6}4s^{2}3d^{5}$, version 02Aug2007), \ch{Co} ($3p^{6}3d^{7}4s^{2}$, version 23Apr2009), \ch{Zn} ($3d^{10}4s^{2}$, version 06Sep2000), \ch{Ni} ($3p^{6}3d^{3}4s^{2}$, version 06Sep2000), and \ch{F} ($2s^{2}2p^{5}$, version 08Apr2002) have been applied.
The $k$-point mesh of Brillouin zone sampling in primitive cell Monkhorst-Pack was set to $8\times8\times8$.
The exchange correlation was represented within the generalized gradient approximation (GGA) in the PBEsol parametrization~\cite{perdew2008restoring}.
Additionally, the $d$ electrons were corrected through the DFT$+U$ ($U=4$\,eV) approximation within the Liechtenstein formalism~\cite{liechtenstein1995density}.  
Born effective charges, dielectric properties, and lattice dynamics were calculated within the density functional perturbation theory (DFPT)~\cite{gonze1997dynamical} as implemented in the \textsc{vasp} code and analyzed through the Phonopy interface~\cite{togo2015first}.
The longitudinal-transverse optical phonon (LO-TO) splitting near the $\Gamma$ point of the Brillouin zone was included using non-analytical corrections to the dynamical matrix~\cite{wang2010mixed}.
Finally, the force constants $k$ of phonon modes were calculated as eigenvalues of the force constant matrix~\footnote{The force constant matrix is defined by $C_{\alpha i, \beta j} = \cfrac{\partial F_{\alpha i}}{\partial r_{\beta j}}$, where $\alpha$ and $\beta$ label the ions, $i$ and $j$ Cartesian directions, $F$ is the force on the ion and $r$ is the ion position, where the acoustic sum rule is imposed to guarantee translation invariance as implemented in Phonopy~\cite{togo2015first}}.

\section{Results and Discussion}
\subsection{Far-infrared spectroscopy}

% \begin{figure*}
% \centering
% \includegraphics[width=2\columnwidth]{fig_reflectivity_20072021_compressed.pdf}
% %\includegraphics[width=2\columnwidth]{fig_reflectivity_13072021_2.pdf}
% \caption{\label{fig:reflectivity}
% (a)--(d) Room temperature and (e)--(h) temperature colormap of far-infrared reflectivity spectra of the cubic fluoroperovskites \ch{KZnF3}, \ch{RbMnF3}, \ch{KNiF3} and \ch{KMgF3}, respectively.
% The solid black lines are results of fits based on the generalized oscillator model according to Eq~\eqref{eq:epsilon_TOLO}.
% Color vertical dashed lines indicate $\omega_{\textrm{TO}}$ and $\omega_{\textrm{LO}}$ phonon frequencies, with $\omega_{\textrm{TO}}<\omega_{\textrm{LO}}$.
% Horizontal white dashed lines indicate antiferromagnetic phase transition temperatures $T_{N}$.
% }
% \end{figure*}

The group-theoretical analysis for the cubic fluoroperovskites predicts five triply degenerate phonons \mbox{$\Gamma_{\textrm{total}} = 4 T_{1u} \oplus T_{2u}$} among which \mbox{$\Gamma_{\textrm{IR}} = 3 T_{1u}$} are IR-active or polar~\cite{kroumova2003bilbao}.
The three reflection bands observed in the far-infrared spectra $R(\omega)$ at room temperature shown in Figs.~\ref{fig:reflectivity}(a)--\ref{fig:reflectivity}(d) originate from the IR-active phonons in the studied crystals. % IR active phonons are related to the three reflection bands observed in the far-infrared spectra $R(\omega)$ at room temperature in the studied crystals as seen in Figs.~\ref{fig:reflectivity}(a)--\ref{fig:reflectivity}(d).
The reflection band widths correspond to the difference between LO and TO frequencies of polar phonons arising from Coulomb interaction.
The far-infrared reflectivity spectra $R(\omega)$ were fitted using the Fresnel equation~\cite{born2013principles}
\begin{equation}
\label{eq:reflectivity}
R(\omega) = \Bigl|\frac{\sqrt{\varepsilon(\omega)} - 1}{\sqrt{\varepsilon(\omega)} + 1}\Bigr|^2,
\end{equation}
with a factorized complex dielectric permittivity~\cite{gervais1974anharmonicity}
\begin{equation}
\label{eq:epsilon_TOLO}
\varepsilon(\omega) = \varepsilon_{1}(\omega) - i\varepsilon_{2}(\omega) = \varepsilon_{\infty}\prod\limits_{j}\frac{{\omega^{2}_{j\textrm{LO}}} - {\omega}^2 + i\gamma_{j\textrm{LO}}\omega}{{\omega^{2}_{j\textrm{TO}}} - {\omega}^2 + i\gamma_{j\textrm{TO}}\omega},
\end{equation}
where $\varepsilon_{\infty}$ is the high-frequency dielectric permittivity, $\omega_{j\textrm{LO}}$, $\omega_{j\textrm{TO}}$, $\gamma_{j\textrm{LO}}$ and $\gamma_{j\textrm{TO}}$ correspond to $\textrm{LO}$ and $\textrm{TO}$ frequencies ($\omega_{j}$) and dampings ($\gamma_{j}$) of the $j$th IR-active phonon, respectively.
There is a good agreement between experimental (green) and fitted (black) lines as shown in Figs.~\ref{fig:reflectivity}(a)--\ref{fig:reflectivity}(d).
Deviations appear only at the highest-frequency phonon presumably due to multiphonon processes involving zone boundary phonons~\cite{young1969temperature}.
The obtained from the fit room temperature values of frequencies $\omega_{j}$ and dampings $\gamma_{j}$ of the $j=1-3$ polar phonons and high-frequency dielectric permittivity $\varepsilon_{\infty}$ are listed in Table~\ref{tab:phonon_parameters}.
These parameters are in satisfactory agreement with the data for room temperature from the literature for the cubic fluoroperovskites~\cite{axe1967infrared,perry1967infrared,balkanski1967infrared,hofmeister1991comparison}.

\begin{figure*}[t]
\centering
\includegraphics[width=2\columnwidth]{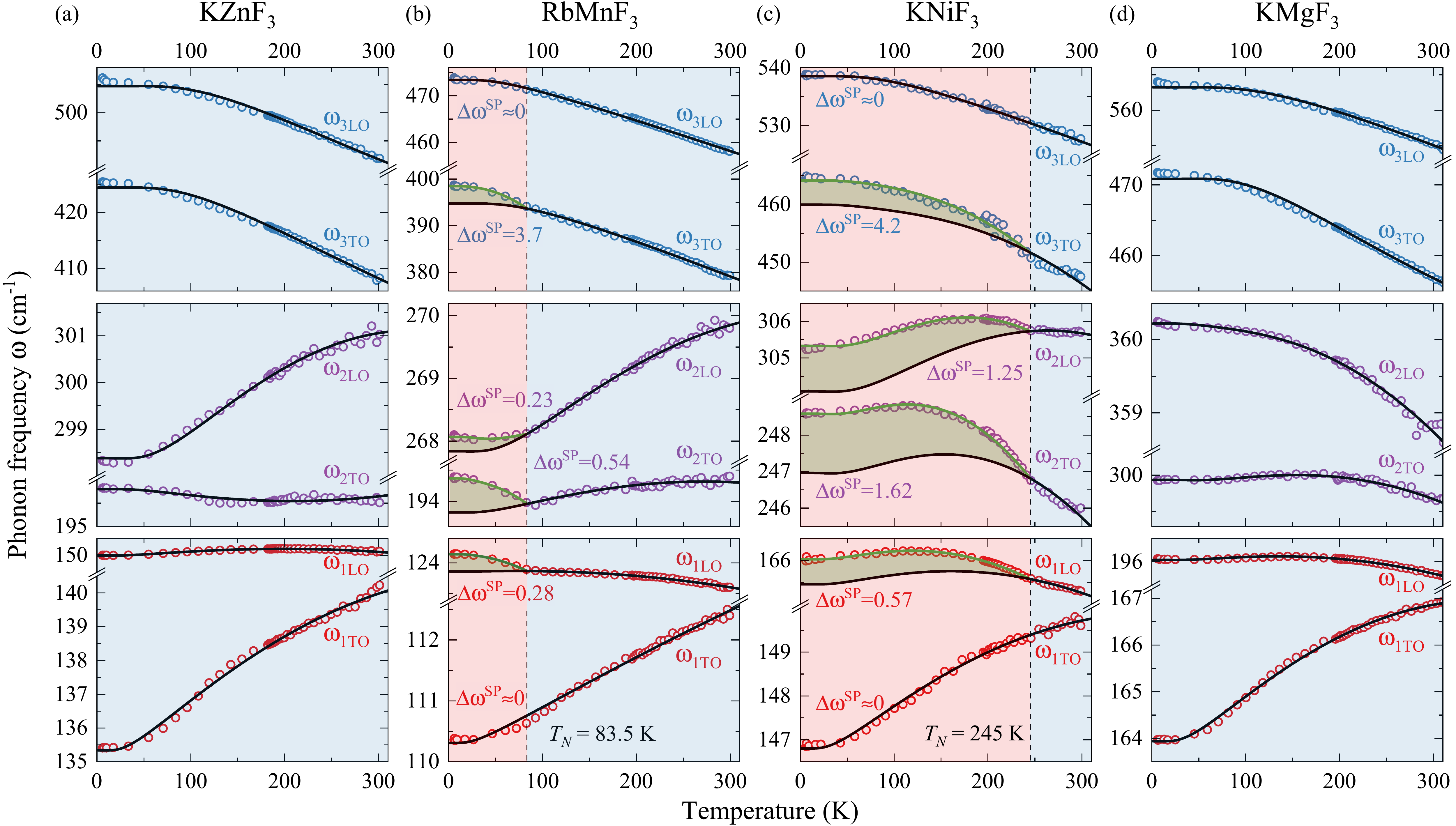}
\caption{\label{fig:phonon}
Temperature dependences of frequencies $\omega_{j}$ of the $j=1-3$ polar phonons for the cubic fluoroperovskites (a)~\ch{KZnF3}, (b)~\ch{RbMnF3}, (c)~\ch{KNiF3}, and (d)~\ch{KMgF3}.
The color circles represent the experimental data.
The black lines correspond to the fit under assumption of anharmonic temperature behavior in the absence of magnetic ordering according to Eq.~\eqref{eq:omega_anharmonism}.
The green lines are fits of the shift due to spin-phonon coupling according to Eq.~\eqref{eq:spin_phonon}.
Values of the spin-phonon coupling $\Delta\omega^{\textrm{SP}}$ (cm$^{-1}$) are given.
Differences between nonmagnetic and magnetic fit functions are shown by green filled areas.
The paramagnetic and antiferromagnetic phases are shown in blue and red color filled backgrounds, respectively.
}
\end{figure*}

For determining the temperature evolution of polar phonons in the studied crystals, we examined the far-infrared reflectivity spectra in the range from 5 to 300\,K which are shown by the colormaps in Figs.~\ref{fig:reflectivity}(e)--\ref{fig:reflectivity}(h).
The fitting of these spectra using Eqs.~\eqref{eq:reflectivity} and~\eqref{eq:epsilon_TOLO} allowed us to obtain the temperature dependences of the $\omega_{j\textrm{TO}}$ and $\omega_{j\textrm{LO}}$ phonon frequencies which are shown by color circles in Fig.~\ref{fig:phonon}.

\subsection{Softening of the polar mode}
% Our analysis of the temperature dependences of the polar phonons frequencies made it possible to reveal an interesting result. 
Our analysis reveals that the phonon frequency $\omega_{1\textrm{TO}}$ decreases (\textit{i.e.} softens), by a few cm$^{-1}$, at cooling in all studied cubic fluoroperovskites as shown in the bottom frames of Fig.~\ref{fig:phonon}.
This softening is similar to that observed in the isostructural \ch{KCoF3} and \ch{RbCoF3} crystals previously reported in Ref.~\cite{dubrovin2019lattice}. 
For our further analysis, it is convenient to convert phonon frequency $\omega$ to the force constant $k$, which are related as $\omega = \sqrt{{k}/{\mu}}$, where $\mu$ is the reduced mass of ions in the unit cell.
According to the general principles of lattice dynamics theory~\cite{born1954dynamical}, the force constant $k$ can be represented as $k = k_{0} + k_{\textrm{ah}}$, where $k_{0}$ is a temperature independent harmonic and $k_{\textrm{ah}}$ is a temperature dependent anharmonic force constant.
It was shown in Ref.~\cite{ridou1984anharmonicity} that in the cubic fluoroperovskites the value of quasi-harmonic force constant $k_{\textrm{qh}}$ related to the thermal expansion of crystal is one order of magnitude lower than the anharmonic force constant $k_{\textrm{ah}}$ that allows us to neglect $k_{\textrm{qh}}$ in the analysis.
Moreover, we suppose that the phonon frequency at the low temperature $\omega_{1\textrm{TO}}(5\,\mathrm{K})$ is determined by the harmonic force constant $k_{0}$ only whereas the anharmonic force constant $k_{\textrm{ah}}$ is neglected.

\begin{figure*}[t]
\centering
\includegraphics[width=2\columnwidth]{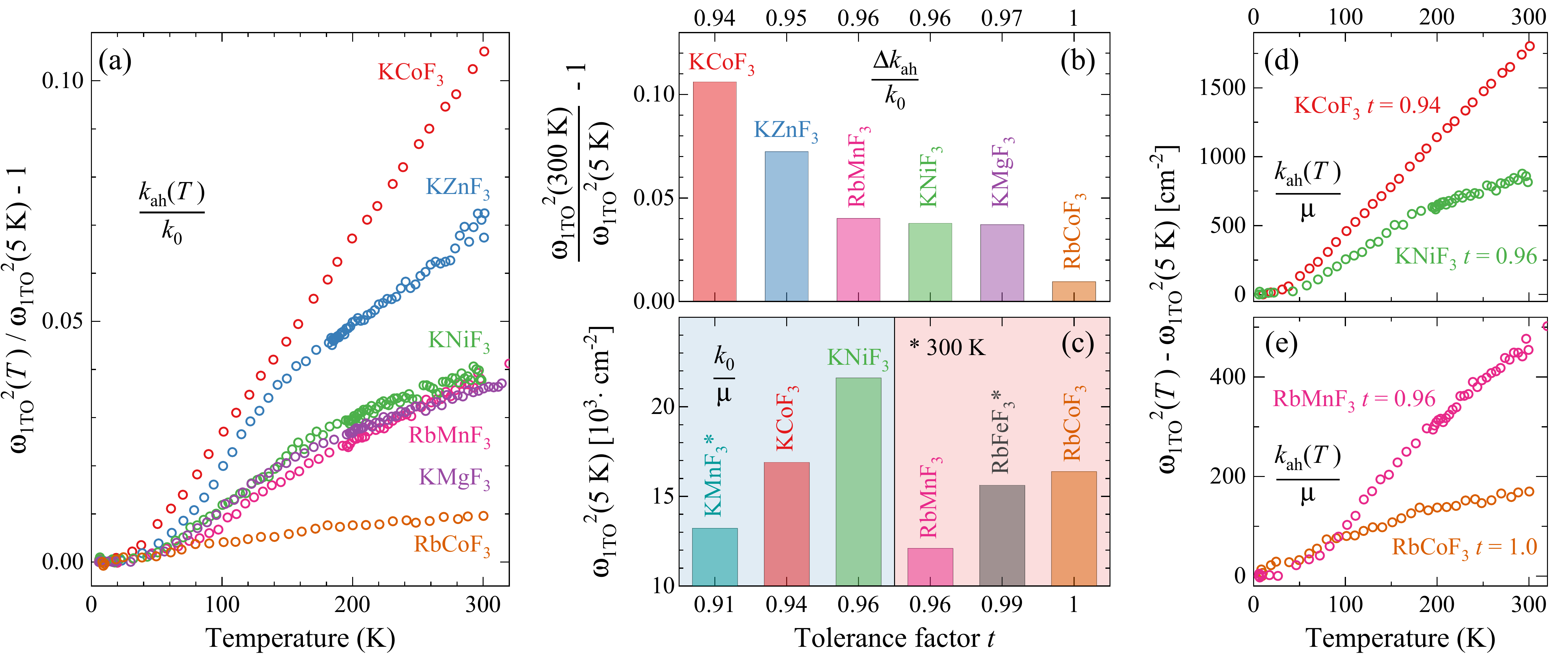}
\caption{\label{fig:soft_mode}
(a)~Temperature and (b)~tolerance factor $t$ (at room temperature) dependences of the differences of squared phonon frequencies $\omega_{1\textrm{TO}}^{2}(T) - \omega_{1\textrm{TO}}^{2}(5\,\textrm{K})$ which represents the ratio of anharmonic and harmonic force constants ${\Delta k_{\textrm{ah}}}/{k_{0}}$ in the cubic fluoroperovskites.
The data for \ch{KCoF3} and \ch{RbCoF3} have been adapted from Ref.~\cite{dubrovin2019lattice}.
(c)~Dependence of the squared phonon frequency $\omega_{1\textrm{TO}}^{2}(5\,\textrm{K})$ at low temperature which represent the reduced harmonic force constant  ${k_{0}(T)}/{\mu}$ on the tolerance factor $t$ in two groups of the cubic fluoroperovskites: \ch{KMnF3}~\cite{axe1967infrared,perry1967infrared} (at room temperature), \ch{KCoF3}~\cite{dubrovin2019lattice}, \ch{KNiF3} and \ch{RbMnF3}, \ch{RbFeF3}~\cite{nakagawa1973transverse} (at room temperature), \ch{RbCoF3}~\cite{dubrovin2019lattice} with insignificant difference of the reduced mass $\mu$.  
Temperature dependences of the $\omega_{1\textrm{TO}}^{2}(T) - \omega_{1\textrm{TO}}^{2}(5\,\textrm{K})$ which represent the reduced anharmonic force constant ${k_{\textrm{ah}}(T)}/{\mu}$ in groups of the cubic fluoroperovskites (d)~\ch{KCoF3}, \ch{KNiF3} and (e)~\ch{RbMnF3}, \ch{RbCoF3} in which the difference in the value of $\mu$ is neglected. 
}
\end{figure*}

% The temperature dependences of the quantity ${\omega_{1\textrm{TO}}^{2}(T)}/{\omega_{1\textrm{TO}}^{2}(5\,\textrm{K})} - 1$ which represents a ratio of anharmonic to harmonic force constants ${k_{\textrm{ah}}(T)}/{k_{0}}$ of the 1TO polar phonon for the cubic fluoroperovskites under study together with the data for \ch{KCoF3} and \ch{RCoF3} from Ref.~\cite{dubrovin2019lattice} are shown in Fig.~\ref{fig:soft_mode}(a).
It is easy to see that the quantity ${\omega_{1\textrm{TO}}^{2}(T)}/{\omega_{1\textrm{TO}}^{2}(5\,\textrm{K})} - 1$ represents a ratio of anharmonic to harmonic force constants ${k_{\textrm{ah}}(T)}/{k_{0}}$ of the 1TO polar phonon.
The temperature dependences of this quantity for the considered cubic fluoroperovskites together with the data for \ch{KCoF3} and \ch{RbCoF3} from Ref.~\cite{dubrovin2019lattice} are shown in Fig.~\ref{fig:soft_mode}(a).
It can be seen that a consistent decrease of the ratio ${k_{\textrm{ah}}(T)}/{k_{0}}$ takes place at cooling, with more pronounced change of this value in \ch{KCoF3} with the lowest tolerance factor $t=0.94$, whereas the smallest reduction was observed in \ch{RbCoF3} with the largest value of $t=1$ among the presented cubic fluoroperovskites.
Figure~\ref{fig:soft_mode}(b) shows the extracted ratios of the force constants at room temperature ${\Delta k_{\textrm{ah}}}/{k_{0}}$ in the studied crystals as a function of the tolerance factor $t$.
%The ratio of the force constants at the room temperature ${\Delta k_{\textrm{ah}}}/{k_{0}}$ in studied crystals depending on the tolerance factor $t$ is shown as a bar graph in Fig.~\ref{fig:soft_mode}(b).
A strong correlation between the ratio ${\Delta k_{\textrm{ah}}}/{k_{0}}$ and the tolerance factor $t$ is observed, while the smaller is the value of $t$, the larger is the ratio of ${\Delta k_{\textrm{ah}}}/{k_{0}}$ in the presented cubic fluoroperovskites.
%It is clearly seen a strong correlation between the ratio ${\Delta k_{\textrm{ah}}}/{k_{0}}$ and the tolerance factor $t$, while the smaller is the value of $t$, the larger is the ratio of ${\Delta k_{\textrm{ah}}}/{k_{0}}$ in the presented cubic fluoroperovskites.

To analyze the behavior of the harmonic force constant $k_{0}$ let us consider the squared frequency $\omega_{1\textrm{TO}}^{2}(5\,\textrm{K})$ at the low temperature also in the two groups of the cubic fluoroperovskites \ch{KMnF3}~\cite{axe1967infrared,perry1967infrared}, \ch{KCoF3}~\cite{dubrovin2019lattice}, \ch{KNiF3} and \ch{RbMnF3}, \ch{RbFeF3}~\cite{nakagawa1973transverse}, \ch{RbCoF3}~\cite{dubrovin2019lattice} in which also only the $3d$ ion changes.
The $k_{0}$ values for \ch{KMnF3} and \ch{RbFeF3} are somewhat overestimated because the phonon frequencies $\omega_{1\textrm{TO}}$ are given in the literature only for room temperature but this does not violate the general trend.
As can be seen in Fig.~\ref{fig:soft_mode}(c), the value $\omega_{1\textrm{TO}}^{2}(5\,\textrm{K})$ is reduced with decrease of the tolerance factor $t$ in each of these two crystal groups, which reflects the corresponding change of the reduced harmonic force constant ${k_{0}}/{\mu}$.
Therefore, this analysis leads us to conclude that in the cubic fluoroperovskites \ch{AMF3} there is a tangible coupling between the tolerance factor $t$ and harmonic force constant $k_{0}$ of the 1TO phonon at which a decrease of the $t$ drives to a clear reduction of the $k_{0}$.

In order to reveal the temperature behavior of the anharmonic force constant $k_{\textrm{ah}}(T)$, the difference of squared phonon frequencies $\omega_{1\textrm{TO}}^{2}(T) - \omega_{1\textrm{TO}}^{2}(5\,\textrm{K})$ which represents the value ${k_{\textrm{ah}}(T)}/{\mu}$ were analyzed in the two groups of the cubic fluoroperovskites \ch{KCoF3}, \ch{KNiF3} and \ch{RbMnF3}, \ch{RbCoF3}, as shown in Figs.~\ref{fig:soft_mode}(d) and~\ref{fig:soft_mode}(e), respectively.
In each of these groups, the difference in values of $\mu$ can be neglected since only the \ch{M} ion belonging to the $3d$ ions changes, which allows comparing the quantity $k_{\textrm{ah}}(T)$ for crystals with different tolerance factor $t$.
According to Figs.~\ref{fig:soft_mode}(d) and~\ref{fig:soft_mode}(e), the value of $\omega_{1\textrm{TO}}^{2}(T) - \omega_{1\textrm{TO}}^{2}(5\,\textrm{K})$ for \ch{KCoF3} ($t=0.94$) and \ch{RbMnF3} ($t=0.96$) is respectively more than for \ch{KNiF3} ($t=0.96$) and \ch{RbCoF3} ($t=1.0$), thus anharmonic force constant $k_{\textrm{ah}}$ increases when the tolerance factor $t$ is decreased in the cubic fluoroperovskites.

% To analyze the behavior of the harmonic force constant $k_{0}$, let us consider the squared frequency $\omega_{1\textrm{TO}}^{2}(5\,\textrm{K})$ at the low temperature also in two groups of cubic fluoroperovskites \ch{KMnF3}~\cite{axe1967infrared,perry1967infrared} (data at room temperature), \ch{KCoF3}~\cite{dubrovin2019lattice}, \ch{KNiF3} and \ch{RbMnF3}, \ch{RbFeF3}~\cite{nakagawa1973transverse} (data at room temperature), \ch{RbCoF3}~\cite{dubrovin2019lattice} in which also only the $3d$ ion changes. 
% As can be seen in Fig.~\ref{fig:soft_mode}(c) the value $\omega_{1\textrm{TO}}^{2}(5\,\textrm{K})$ reduces with decrease of the tolerance factor $t$ in each of these two groups of crystals, which reflects the corresponding change of the reduced harmonic force constant ${k_{0}}/{\mu}$.
% Thus, this allows us to conclude that in cubic fluoroperovskites \ch{AMF3} there is a coupling between the tolerance factor $t$ and harmonic force constant $k_{0}$ of the 1TO mode in which a decrease of the $t$ leads to reduce of the $k_{0}$.

\begin{figure*}[t]
\centering
\includegraphics[width=2\columnwidth]{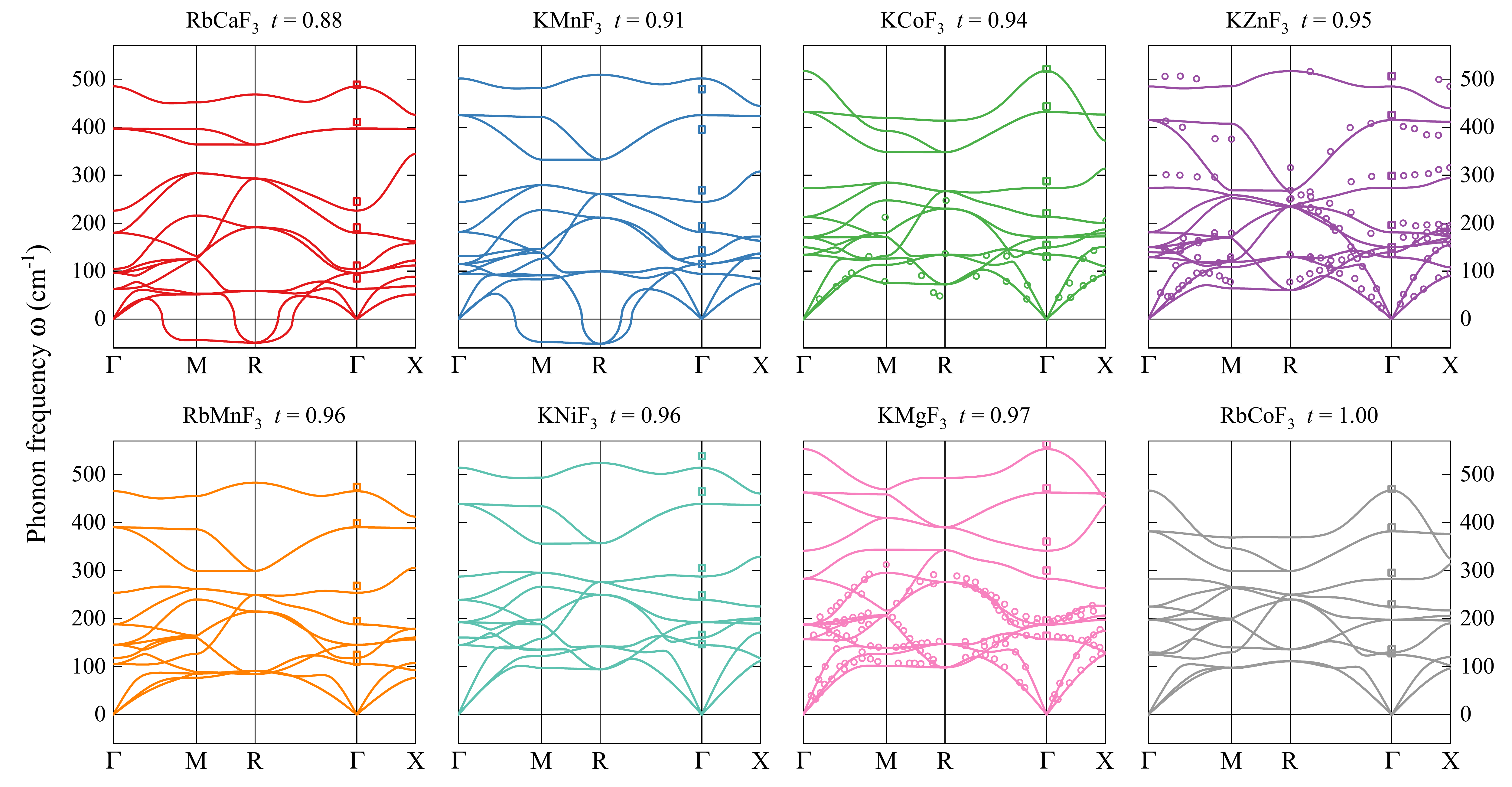}
\caption{\label{fig:dft_phonon_spectra}
Phonon dispersion curves along the $\Gamma$ -- $M$ -- $R$ -- $\Gamma$ -- $X$ high-symmetry path of the Brillouin zone of the cubic fluoroperovskites with different tolerance factors $t$.
\textcolor{newtext}{
The lines represent the calculated data.
Imaginary frequencies are signified by negative numbers and correspond to unstable phonons.
The circles denote the adapted experimental data from the neutron scattering for \ch{KCoF3}~\cite{holden1971excitations_1}, \ch{KZnF3}~\cite{lehner1982lattice}, and \ch{KMgF3}~\cite{salatin1993lattice}.
The squares represent the polar phonon frequencies in the Brillouin zone center from our infrared spectroscopy experiments for \ch{KZnF3}, \ch{RbMnF3}, \ch{KNiF3}, and \ch{KMgF3}, and adapted from the literature for \ch{RbCaF3}~\cite{ridou1986temperature}, \ch{KMnF3}~\cite{axe1967infrared}, \ch{KCoF3}~\cite{dubrovin2019lattice}, and \ch{RbCoF3}~\cite{dubrovin2019lattice}.}}
\end{figure*}

A similar behavior of the polar phonons, but with more pronounced temperature changes, was observed in the incipient ferroelectrics such as oxide perovskites \ch{SrTiO3}~\cite{muller1979srti}, \ch{CaTiO3}~\cite{lemanov1999perovskite} and \ch{EuTiO3}~\cite{kamba2007magnetodielectric}.
In these materials there is a soft polar mode which frequency $\omega_{\textrm{SM}}$ tends to but does not reach zero at cooling up to the lowest temperatures and the expected ferroelectric phase transition does not occur~\cite{kamba2021soft}.
The Curie temperature $T_{C}$ in incipient ferroelectrics at which the frequency $\omega_{\textrm{SM}}$ or its extrapolated linear part goes to zero is usually negative, but sometimes it is positive what suggests a ferroelectric transition which nevertheless is not realized due to a rather low value of this temperature~\cite{kvyatkovskii2001quantum}.
Moreover, the harmonic force constant $k_{0}$ of the soft polar mode in incipient ferroelectrics has a small absolute value that is negative, similar to the case of ferroelectrics, or positive~\cite{kvyatkovskii2001quantum}.
As mentioned above, the temperature behavior of the phonon frequency $\omega$ is determined by the anharmonic force constant $k_{\textrm{ah}}$, the sign and value of which is defined by the mutual compensation of the two terms with the opposite temperature dependence of the third and fourth order in anharmonicity~\cite{bruce1973lattice,bruce1981structural}.
Thus, the $k_\textrm{ah}$ is positive for ferroelectrics and incipient ferroelectrics but it is negative for normal insulators.
A delicate balance between these anharmonic terms occurs in the cubic fluoroperovskites at which anharmonicities mutually compensate each other when $t$ approaches to 1.   
We believe that the revealed correlation between the harmonic $k_{0}$, anharmonic $k_{\textrm{ah}}$ force constants, and their ratio ${k_{\textrm{ah}}(T)}/{k_{0}}$ with the tolerance factor $t$ clearly indicates the existence of incipient ferroelectric instability in the cubic fluoroperovskites.
It is worth noting that a similar consistent increase of the anharmonic force constant $\Delta k_{\textrm{ah}}$ with decreasing of the harmonic force constant $k_{0}$ was previously observed in the group of the IV--VI materials \ch{PbS}, \ch{PbSe}, \ch{PbTe}, and \ch{SnTe}, but according to Ref.~\cite{kvyatkovskiui1988microscopic} this effect cannot be explained within the framework of the phenomenological theory of lattice anharmonicity.

\subsection{Lattice dynamics simulations}
% \begin{figure*}[t]
% \centering
% \includegraphics[width=2\columnwidth]{fig_dft_phonon_spectra_20072021_compressed.pdf}
% %\includegraphics[width=2\columnwidth]{fig_dft_phonon_spectra_30062021.pdf}
% %\includegraphics[width=2\columnwidth]{fig_dft_phonon_spectra_08062021.pdf}
% \caption{\label{fig:dft_phonon_spectra}
% Calculated phonon dispersion curves along the $\Gamma$ -- $M$ -- $R$ -- $\Gamma$ -- $X$ high-symmetry path of the Brillouin zone of the cubic fluoroperovskites with different tolerance factors $t$.
% Imaginary frequencies are represented by negative numbers and correspond to unstable phonons. 
% }
% \end{figure*}

To reveal the features of the lattice dynamics of the cubic fluoroperovskites under study, we performed the calculations of the phonon dispersion curves considering the LO-TO splitting along the $\Gamma$ -- $M$ -- $R$ -- $\Gamma$ -- $X$ high-symmetry path of the Brillouin zone [see Fig.~\ref{fig:structure}(c)].
We also included the \ch{RbCaF3}, \ch{KMnF3}, \ch{KCoF3}, and \ch{RbCoF3} crystals to fully cover the $0.88<t<1.0$ range of stability of the cubic fluoroperovskites as shown in Fig.~\ref{fig:dft_phonon_spectra}.
The G-type antiferromagnetic (G-AFM) spin configuration was considered for the case of magnetic crystals.
The obtained results are in satisfactory agreement with the reported data~\cite{holden1971excitations_1,lehner1982lattice,becher1989simulation,salatin1993lattice,salaun1995determination,vaitheeswaran2016calculated,ehsan2018dft} \textcolor{newtext}{as can be seen in Fig.~\ref{fig:dft_phonon_spectra}.}
The calculated frequencies $\omega$ and force constants $k$ of the low-lying $T_{1u}$, $X_{5}$, $M_{2}$ and $R_{15'}$ phonons at the $\Gamma$, $X$, $M$ and $R$ points of the Brillouin zone, respectively, together with the obtained lattice parameters $a$ and tolerance factors $t$ of the cubic fluoroperovskites are listed in Table~\ref{tab:dft_phonons_BZ}.
For the studied crystals, the obtained phonon frequencies $\omega$, dielectric strengths $\Delta\varepsilon$, and permittivities $\varepsilon_{0}$ and $\varepsilon_{\infty}$ at the $\Gamma$ point, as well as the lattice parameters $a$ are listed in Table~\ref{tab:dft_phonon_parameters}.
It can be noted that the obtained values are in fair agreement to the experimental data presented in Table~\ref{tab:phonon_parameters} \textcolor{newtext}{and in the literature~\cite{perry1967infrared,axe1967infrared,balkanski1967infrared,nakagawa1973transverse,nakagawa1974infrared,ridou1986temperature,dubrovin2019lattice}.}

\begin{figure}
\centering
\includegraphics[width=\columnwidth]{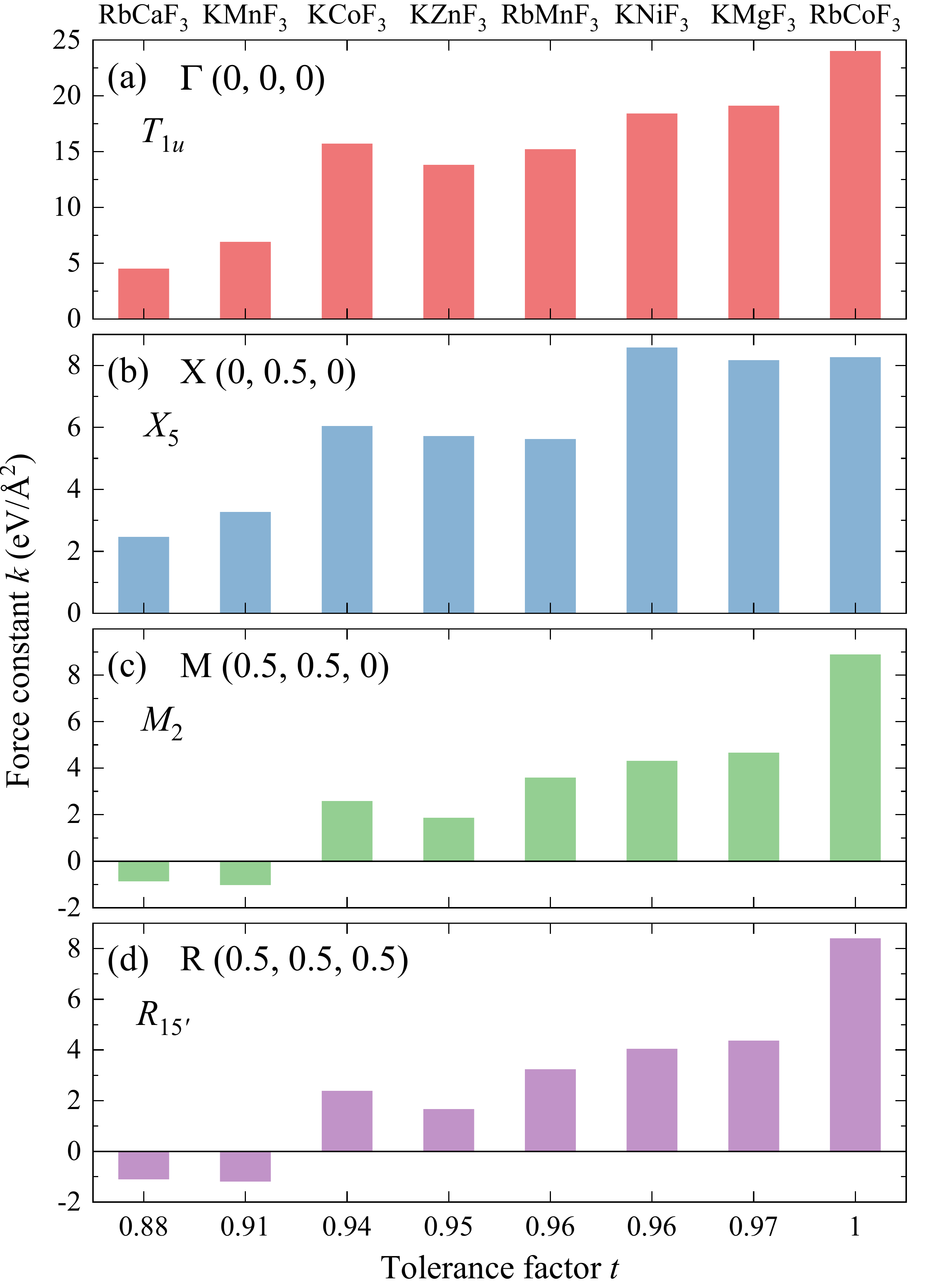}
\caption{\label{fig:dft_DZ_force}
Calculated force constants $k$ of the low-lying $T_{1u}$, $X_{5}$, $M_{2}$, and $R_{15'}$ phonons at the $\Gamma$, $X$, $M$, and $R$ high-symmetry points of the Brillouin zone, respectively, in the cubic fluoroperovskites \ch{AMX3} with different tolerance factors $t$.
}
\end{figure}

The cubic structure is dynamically stable for the vast majority of fluoroperovskites under study.
However, the computed harmonic force constants $k$ of phonons monotonically decrease with lowering of the tolerance factor $t$ at all high-symmetry points of the Brillouin zone, as shown in Fig.~\ref{fig:dft_DZ_force}.
As a result, the crystals with a small tolerance factors \ch{RbCaF3} ($t=0.88$) and \ch{KMnF3} ($t=0.91$) have strongly unstable antiferrodistortive modes $R_{15'}$ and $M_{2}$ with negative force constants $k$ and imaginary frequencies $\omega$ at the $R$ and $M$ high-symmetry points and along the $M$--$R$ symmetry line, as it can be seen from Figs.~\ref{fig:dft_phonon_spectra}, \ref{fig:dft_DZ_force}(c), \ref{fig:dft_DZ_force}(d), and Table~\ref{tab:dft_phonons_BZ}.   
It is interesting to note that in the studied fluoroperovskites the lowest-frequency phonon mode covering a continuum of wave vectors along the $M$--$R$ edge of the Brillouin zone is sufficiently flat as shown in Fig.~\ref{fig:dft_phonon_spectra}, which is similar to the case found for some other cubic perovskites~\cite{lasota1997abinitio,rushchanskii2012first,lanigan2021two}.
The $R_{15'}$ phonon force constant $k$ tending to zero at cooling drives the phase transition from the cubic $Pm\overline{3}m$ to the tetragonal $I4/mcm$ structure which was experimentally observed at $T_{1}=193$\,K in \ch{RbCaF3}~\cite{rousseau1977rbcaf3} and 186\,K in \ch{KMnF3}~\cite{kapusta1999revised}.
A further temperature decrease leads to the zeroing of the $M_{2}$ phonon force constant $k$ which induces the phase transition to the orthorhombic $Pnma$ structure below $T_{2}=65$\,K in \ch{RbCaF3}~\cite{knight2018high} and at 75\,K in \ch{KMnF3}~\cite{knight2020nuclear}.
It should be noted that in \ch{CsCaF3} ($t=0.94$) and \ch{KZnF3} ($t=0.95$) crystals with a slightly higher tolerance factors, the experimental frequency of the low-lying phonon at the $R$ point is also decreased at cooling, but does not reach a zero~\cite{ridou1984anharmonicity,lehner1982lattice}.
Moreover, the experimental phonon frequencies at the $R$ point in the cubic fluoroperovskites from Ref.~\cite{holden1971excitations_1} correspond to the trend shown in Fig.~\ref{fig:dft_DZ_force}(d).
The force constants $k$ at the $\Gamma$ and $X$ points are positive for all cubic fluoroperovskites with $0.88<t<1.0$ and become negative only in the high-symmetry cubic structure of the orthorhombic crystals with low values of $t$~\cite{garcia2014geometric}.
The dependence of the computed harmonic force constant $k$ of the 1TO phonon at the $\Gamma$ point on the tolerance factor $t$ is very close to those obtained for the $k_{0}$ from experiments, as shown in Figs.~\ref{fig:dft_DZ_force}(a) and~\ref{fig:soft_mode}(c), respectively.

Thus, we revealed the correlations between the force constants $k$ values and the tolerance factor $t$, such that with a reduction of the $t$ the values of $k$ are decreased at all high-symmetry points of the Brillouin zone in the cubic fluoroperovskites.
The obtained results are in good agreement with the calculated data for fluoroperovskites in Ref.~\cite{garcia2014geometric}.
Moreover, the computed ionic Born effective charges in the studied crystals are close to the nominal values in contrast to the ferroelectric oxide perovskites~\cite{zhong1994giant} and without any correlations on the tolerance factor $t$ as shown in Table~\ref{tab:dft_born_charges}.
Remarkably, our analysis suggests that the discovered trend in the cubic fluoroperovskites has a geometric origin related to steric effect, namely, by the volume filling of the unit cell by ions with different ionic radii.
This size effect was previously used to explain the multiferroicity observed in the \ch{BaMF4} crystal family~\cite{ederer2006origin,garcia2018direct}.
It is worth noting that no such clear correlation between calculated frequencies $\omega$, force constants $k$, and lattice parameters $a$ was observed in our study,  emphasizing the exceptional importance of the tolerance factor $t$ for fluoroperovskites.

\begin{figure}
\centering
\includegraphics[width=\columnwidth]{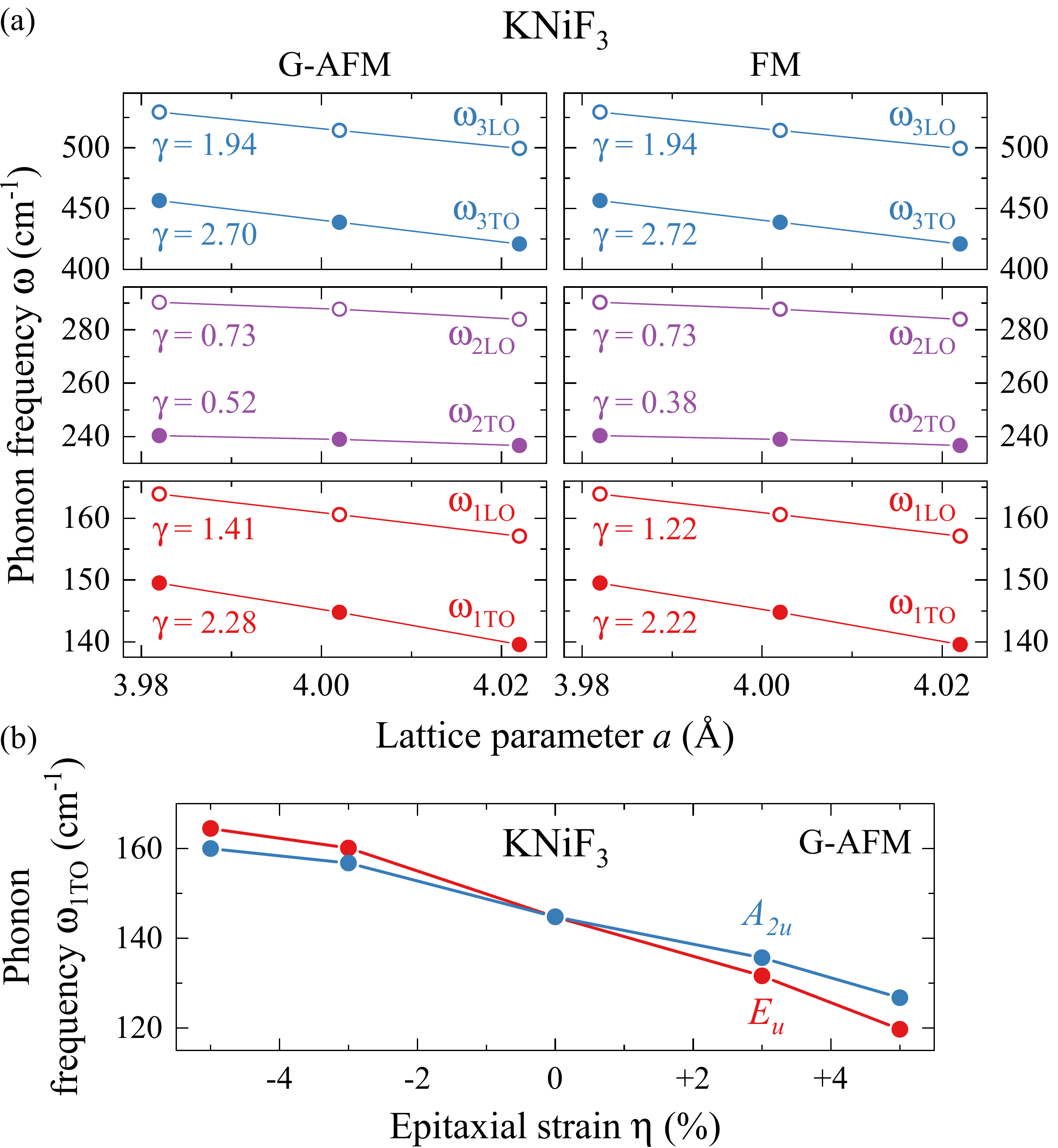}
\caption{\label{fig:KNiF3_volume}
(a)~Calculated phonon frequencies $\omega$ of the polar phonons in cubic fluoroperovskite \ch{KNiF3} in (left frames) G-AFM and (right frames) FM states as a function of lattice parameter $a$.
The equilibrium value of $a$ is in the center.
Values of the obtained Gr{\"u}neisen parameters $\gamma$ for phonons are given.
(b)~Phonon frequency $\omega_{1\textrm{TO}}$ for \ch{KNiF3} in G-AFM state as a function of biaxial epitaxial strain $\eta$.
Negative and positive signs of $\eta$ corresponds to compression and expansion, respectively.
For $\eta\neq{0}$ the symmetry is tetragonal with the space group $P4/mmm$.
}
\end{figure}

In order to gain insight into the quasi-harmonic behavior of lattice dynamics of the cubic fluoroperovskites, we calculated the frequencies $\omega$ of polar phonons at lattice parameters $a$ change on $\pm0.5\%$ with respect to the equilibrium in \ch{KNiF3}.
We assume that the volume trend of phonons of the other cubic fluoroperovskites is similar to the \ch{KNiF3} due to the closeness of their dynamical properties.
Figure~\ref{fig:KNiF3_volume}(a) shows that the $\omega$ of all phonons increase (\textit{i.e.} hardens) with decreasing of $a$ in the both G-AFM and FM (ferromagnetic) states.
Thus, the observed softening of the low-frequency 1TO polar phonons cannot be explained by the quasi-harmonic contribution from thermal contraction of crystals at cooling and it is caused by the anharmonic effects.
The obtained values of the Gr{\"u}neisen parameters $\gamma$ for polar phonons, which quantitatively reflect sensitivity of the frequency to the crystal volume changes, are also presented in Fig.~\ref{fig:KNiF3_volume}(a).

It should be noted that there is an essential difference between cubic fluoroperovskites and incipient ferroelectrics, despite the proximity of the anharmonic behavior of the softening polar phonons.
Thus, it was theoretically predicted and experimentally corroborated that incipient ferroelectrics become genuine ones under epitaxial strain, e.g., for \ch{SrTiO3}~\cite{pertsev2000phase,haeni2004room}, \ch{EuTiO3}~\cite{fennie2006magnetic,lee2010strong}, and \ch{NaMnF3}~\cite{garcia2016strain,yang2017room}.
According to our calculations, the compressive and  tensile biaxial epitaxial strain $\eta$ does not lead the phonon frequency $\omega_{1\textrm{TO}}$ to cross zero up to $\eta=\pm{5}\%$, as shown in Fig.~\ref{fig:KNiF3_volume}(b).
Therefore, the studied cubic fluoroperovskites are not incipient ferroelectrics \textcolor{newtext}{in contrast to the orthorhombic \ch{NaMnF3}~\cite{garcia2016strain,yang2017room,dubrovin2018unveiling,dubrovin2020incipient}}.
\textcolor{newtext}{Moreover, most cubic fluoroperovskites that are stable at zero pressure experimentally showed no structural changes even under high pressure~\cite{vaitheeswaran2007high,aguado2008high,vaitheeswaran2010high,mishra2011high}.}

\subsection{Spin-phonon coupling}

To reveal the antiferromagnetic ordering effects on the phonon landscape, we fitted the temperature dependences of phonon frequencies in the diamagnetic and paramagnetic phases using the expression~\cite{balkanski1983anharmonic}
\begin{eqnarray}
\label{eq:omega_anharmonism}
\omega_{j}(T) = \omega_{j0} + A_{j} \left( 1 + \frac{2}{e^{\hbar\omega_{j0}/2k_{B}T} - 1} \right)
\nonumber\\
+ B_{j} \left( 1 + \frac{3}{e^{\hbar\omega_{j0}/3k_{B}T} - 1} + \frac{3}{(e^{\hbar\omega_{j0}/3k_{B}T} - 1)^2} \right),
\end{eqnarray}
where $\omega_{j0}$ is the harmonic frequency of the $j$th phonon, $A_{j}$ and $B_{j}$ are parameters describing three- and four-phonon anharmonic processes, respectively.
In this simple model, it is assumed that an optical phonon with frequency $\omega_{j0}$ decays into the two (three-phonon process) and three (four-phonon process) acoustic phonons with frequencies ${\omega_{j0}}/{2}$ and ${\omega_{j0}}/{3}$ satisfying both energy and momentum conservation.
In the real crystals, the phonon anharmonicity is more complicated and contributions from these decays processes are usually very small~\cite{mendez1984temperature,lan2012phonon}.
However, Eq.~\eqref{eq:omega_anharmonism} often gives good fits to experimental data, including the case of the studied fluoroperovskite crystals in the paramagnetic and diamagnetic phases as shown by the black lines in the temperature range denoted by the blue background in Fig.~\ref{fig:phonon}.
Due to oversimplified approximation, the model parameters may be physically incorrect and are not given and discussed.

The deviations of experimental frequencies of the polar phonons from the anharmonic fits below the N{\'e}el temperature $T_{N}$ in the antiferromagnetic \ch{RbMnF3} and \ch{KNiF3} caused by the spin-phonon coupling are shown in Figs.~\ref{fig:phonon}(b) and~\ref{fig:phonon}(c), respectively.
These frequency shifts, which are the result of magnetic ordering, are described by the function~\cite{cottam2019spin}
\begin{equation}
\label{eq:spin_phonon}
\omega^{\textrm{AFM}}(T) = \omega^{\textrm{NM}}(T) + \Delta\omega^{\textrm{SP}} \langle S_{i}\cdot{}S_{j} \rangle,
\end{equation}
where $\omega^{\textrm{AFM}}(T)$ and $\omega^{\textrm{NM}}(T)$ are phonon frequency temperature dependences in the antiferromagnetic (AFM) and in hypothetical non-magnetic (NM) phases, $ \langle S_{i}\cdot{}S_{j} \rangle$ denotes a spin-pair correlation function, and $\Delta\omega^{\textrm{SP}}$ is the spin-phonon coupling constant which is equal to the phonon frequency shift at the low temperature.
The spin-pair correlation function $ \langle S_{i}\cdot{}S_{j} \rangle$, neglecting the short-range magnetic ordering, is proportional to $M^{2}$, where $M$ is the magnetic order parameter, which temperature dependence can be described using the Brillouin function~\cite{darby1967tables}
\begin{equation}
\label{eq:brillouin}
B(x) = \frac{M}{M_{0}} = \frac{2S + 1}{2S} \coth{\Bigl(\frac{2S + 1}{S}x\Bigr)} - \frac{1}{2S} \coth{\Bigl(\frac{x}{2S}\Bigr)},
\end{equation}
where $x = \cfrac{3S}{S + 1} \cfrac{M}{M_{0}} \cfrac{T_{N}}{T}$ and $S$, $T$, $T_{N}$, $M$ and $M_{0}$ are spin value, temperature, N{\'e}el temperature, spontaneous magnetization, and full magnetization, respectively.
The difference between anharmonic fits and experimental data in the antiferromagnetic phases were fitted using Eq.~\eqref{eq:spin_phonon} as shown by the green lines in Figs.~\ref{fig:phonon}(b) and~\ref{fig:phonon}(c) for \ch{RbMnF3} and \ch{KNiF3}, respectively.
Besides, the obtained values of spin-phonon coupling constant $\Delta\omega^{\textrm{SP}}$ for all phonons are also given in Figs.~\ref{fig:phonon}(b) and~\ref{fig:phonon}(c).

It is clearly seen in Figs.~\ref{fig:phonon}(b) and~\ref{fig:phonon}(c) that in both crystals appreciable shifts $\Delta\omega^{\textrm{SP}}$ below $T_{N}$ due to the spin-phonon coupling are exhibited by only $\omega_{\textrm{1LO}}$, $\omega_{\textrm{2TO}}$, $\omega_{\textrm{2LO}}$, and $\omega_{\textrm{3TO}}$ phonon frequencies which is qualitatively similar to the results previously observed in \ch{KCoF3} and \ch{RbCoF3}~\cite{dubrovin2019lattice}.
However, in \ch{RbMnF3} and \ch{KNiF3} the spin-phonon coupling was observed for $\omega_{\textrm{3TO}}$, whereas in \ch{KCoF3} and \ch{RbCoF3} the polar phonon with frequency $\omega_{\textrm{3LO}}$ was susceptible to the magnetic ordering.
The reason for this difference can presumably be related to the influence of the strongly anisotropic \ch{Co^{2+}} ion possessing the largest orbital momentum among the $3d^{n}$ ions, which leads to symmetry lowering at the magnetostructural phase transition in \ch{KCoF3} and \ch{RbCoF3} caused by the spin-orbit interaction~\cite{julliard1975analyse}.
The obtained spin-phonon coupling constants $\Delta\omega^{\textrm{SP}}$ for phonons with $\omega_{\textrm{1LO}}$, $\omega_{\textrm{2TO}}$ and $\omega_{\textrm{2LO}}$ frequencies in \ch{KNiF3} are noticeably larger than in \ch{RbMnF3} presumably due to the difference in the values of exchange integrals which appears in the corresponding distinction of $T_{N}$ in these crystals~\cite{barocchi1978determination,chinn1970two}.   
It is known that the $\Delta\omega^{\textrm{SP}}$ is proportional to the second derivative of the exchange integral concerning the ion displacements for phonon~\cite{granado1999magnetic}. 
Moreover, the mass of the \ch{Rb^{1+}} ion (85.5) significantly exceeds that of \ch{K^{1+}} (39.1), which also leads to the value of $\Delta\omega^{\textrm{SP}}$ in \ch{KNiF3} is more than in \ch{RbMnF3}~\cite{granado1999magnetic}.
The phonon with $\omega_{3\textrm{TO}}$ frequency has a relatively large value of $\Delta\varepsilon^{\textrm{SP}}$ which differs slightly in \ch{RbMnF3} and \ch{KNiF3} supposedly related to its strong anharmonicity, which leads to large temperature changes as shown in Figs.~\ref{fig:phonon}(b) and~\ref{fig:phonon}(c), respectively.

\begin{figure*}
\centering
\includegraphics[width=2\columnwidth]{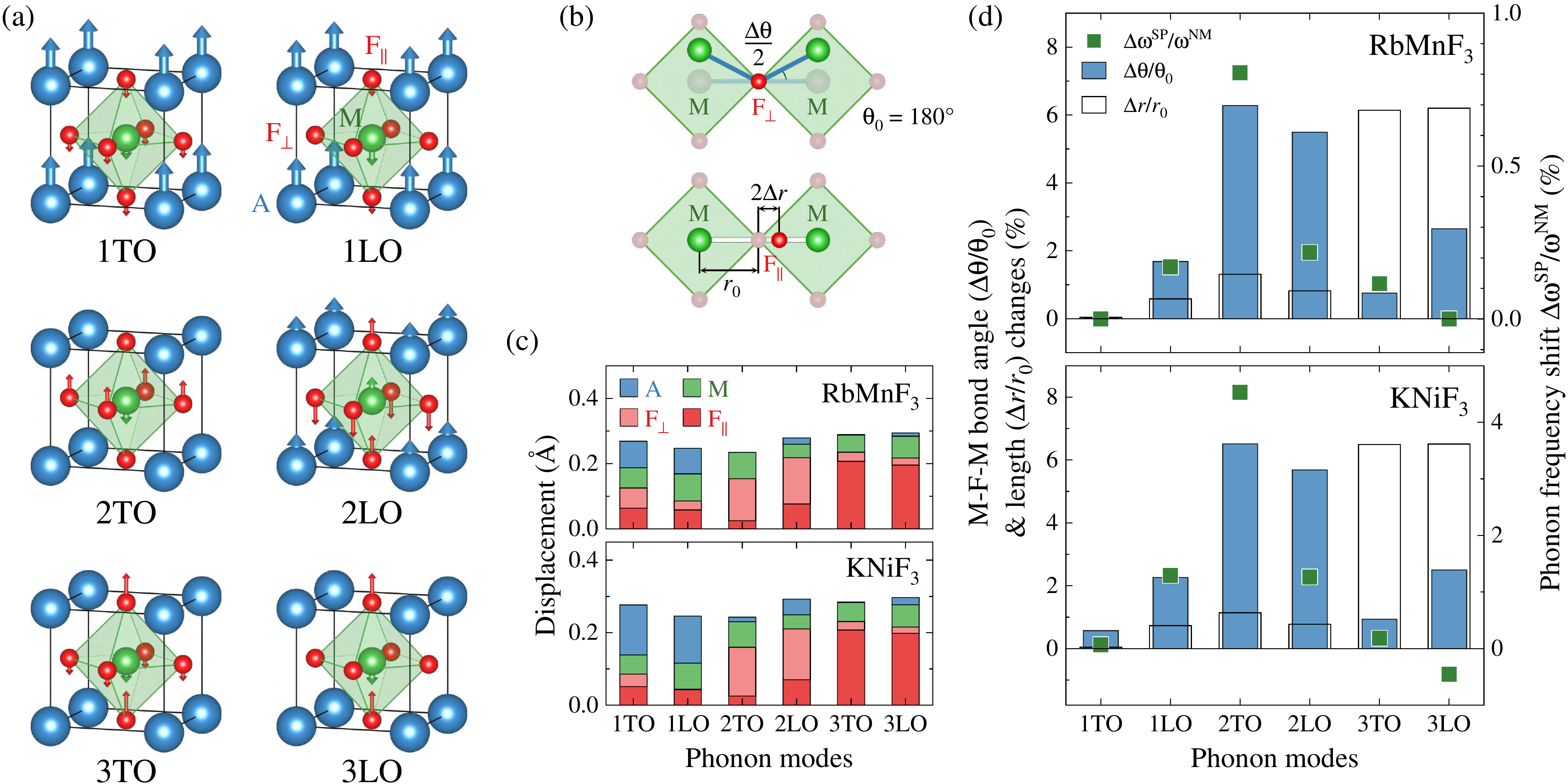}
\caption{\label{fig:spin_phonon}
(a)~Sketch of the ion displacements for polar phonons in the cubic fluoroperovskites according to the DFT simulations. % Arrows indicate the ion displacements.
(b)~The \ch{M}--\ch{F}--\ch{M} superexchange bond angle $\theta_{0}$ and length $r_{0}$ which are dynamically changed due to ion displacements of polar phonons.
$\textrm{F}_{\perp}$ and $\textrm{F}_{\parallel}$ indicate the displacements perpendicular and parallel to the \ch{M}--\ch{F}--\ch{M} bond, respectively.
(c)~Computed absolute values of ion displacements related to polar phonons in \ch{RbMnF3} and \ch{KNiF3}.
(d)~Relationship between the calculated dynamical changes of the bond angle ${\Delta\theta}/{\theta_{0}}$ and length ${\Delta{r}}/{r_{0}}$ and the computed frequency shift ${\Delta\omega^{\textrm{SP}}}/{\omega^{\textrm{NM}}}$ due to the spin-phonon coupling in \ch{RbMnF3} and \ch{KNiF3}.
Picture was prepared using the \textsc{VESTA} software~\cite{momma2011vesta}.
}
\end{figure*}

To grasp further insights on the origin of observed spin-phonon coupling in the cubic fluoroperovskites, we performed lattice dynamics DFT simulations for \ch{RbMnF3} and \ch{KNiF3} also in FM spin configuration.
Since the NM phase is not accessible for the DFT calculations of the magnetic crystals, we assumed for it the averaged values between G-AFM and FM states according to Ref.~\cite{schleck2010elastic}.
For studied magnetic crystals, the computed frequencies $\omega$ of polar phonons at the $\Gamma$ point for G-AFM, FM, and NM states together with values of spin-phonon coupling shifts $\Delta\omega^{\textrm{SP}}=\omega^{\textrm{AFM}}-\omega^{\textrm{NM}}$ are listed in Table~\ref{tab:dft_SP}.
There is a good qualitative agreement between experimental and computed $\Delta\omega^{\textrm{SP}}$ in \ch{RbMnF3} and \ch{KNiF3} as it can be seen from Figs.~\ref{fig:phonon}(b)--\ref{fig:phonon}(c) and Table~\ref{tab:dft_SP}.
The calculated frequency shift $\Delta\omega^{\textrm{SP}}$ due to spin-phonon coupling of the 1TO phonon has a meager absolute value in both crystals, in full agreement with our experiments.
Furthermore, according to our calculations, the 1LO, 2TO, and 2LO phonons have a positive sign of $\Delta\omega^{\textrm{SP}}$ with a most pronounced effect for the 2TO phonon which corresponds to that observed in experiments.
The 3TO phonon exhibits appreciable positive frequency shift $\Delta\omega^{\textrm{SP}}$ in simulations which also agrees with our experimental results.
However, according to Figs.~\ref{fig:phonon}(b) and~\ref{fig:phonon}(c), the spin-phonon coupling for the polar phonons with frequency $\omega_{3\textrm{LO}}$ is insignificant in both crystals which is in agreement with the calculation results for \ch{RbMnF3} only, while for \ch{KNiF3} the $\Delta\omega^{\textrm{SP}}$ has a significant negative value as it can be observed in Table~\ref{tab:dft_SP}.

Figure~\ref{fig:spin_phonon}(a) shows the computed ion displacements for polar phonons in the considered cubic fluoroperovskites.
The TO polar phonon displacements are close to previously published data these crystals~\cite{harada1970determination,nakagawa1973transverse}.
The lowest-frequency 1TO phonon corresponds to the so called Last mode with opposite vibration of the \ch{A} cations and the \ch{MF6} octahedra.
The second 2TO phonon corresponds to the vibrations of the \ch{M} cations against the fluoride octahedra which are known as the Slater mode.
Note that the Slater mode has the lowest frequency in the lead-free oxide perovskites, e.g., in \ch{KNbO3}, \ch{BaTiO3}, \ch{SrTiO3} and \ch{EuTiO3}, and this mode dominates in the ferroelectrics and incipient ferroelectrics~\cite{harada1970determination,hlinka2006infrared,rushchanskii2012first}.
The 3TO polar phonon with the highest frequency represents the bending of the \ch{MF6} octahedra, which corresponds to the Axe mode.
The ion displacements for the LO polar phonons in the cubic fluoroperovskites are also shown in Fig.~\ref{fig:spin_phonon}(a). 

The spin-phonon coupling originates from the dependence of the exchange interaction on the positions of ions.
In the antiferromagnetic cubic fluoroperovskites, phonons dynamically change the bond angle $\theta_{0}$ and length $r_{0}$ of the superexchange \ch{M}--\ch{F}--\ch{M} pathway as shown in Fig.~\ref{fig:spin_phonon}(b).
According to the Goodenough-Kanamori rules, the superexchange interaction is given by a relation $J \propto \cfrac{\tau^{2}}{\Delta - J_{\textrm{H}}}$, where $\tau$ is the hopping integral which is proportional to the effective overlap between the wave functions of electron orbitals of the magnetic \ch{M} cations via the \ch{F} anion, $\Delta$ is the energy difference between the \ch{F} and \ch{M} ion orbitals, and $J_{\textrm{H}}$ is the Hund coupling~\cite{goodenough1963magnetism,lee2011large}.
The superexchange bond angle variation $\Delta\theta$ by phonons affects to the exchange interaction $J$ through the change of the hopping integral $\tau$, whereas the length variation $\Delta{r}$ leads to a change of the energy difference between orbitals $\Delta$.
The calculated absolute values of ion displacements for the TO and LO polar phonons in \ch{RbMnF3} and \ch{KNiF3} are presented as stacked bar graph in Fig.~\ref{fig:spin_phonon}(c).
Therefore, this allows us to reveal the origin of the spin-phonon coupling observed in studied antiferromagnetic crystals.

\begin{figure*}[t]
\centering
\includegraphics[width=2\columnwidth]{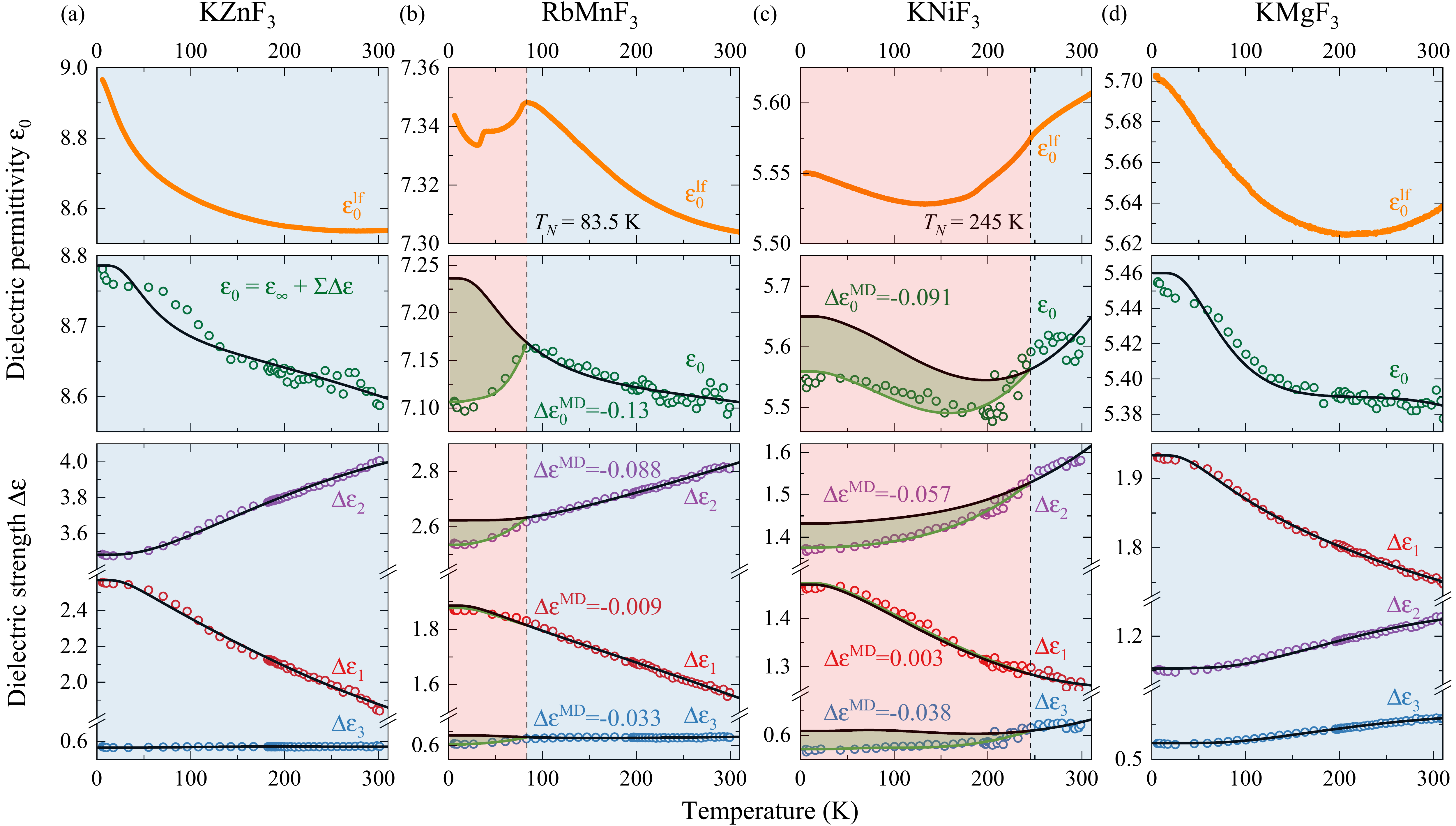}
\caption{\label{fig:dielectric}
Temperature dependences of (bottom frames) the dielectric strength $\Delta\varepsilon_{j}$ for $T_{1u}$ phonons $j=1-3$, (middle frames) the static dielectric permittivity $\varepsilon_{0}$, and (upper frames) the low-frequency dielectric permittivity $\varepsilon^{\textrm{lf}}_{0}$ at $f=100$\,kHz for the cubic fluoroperovskites (a)~\ch{KZnF3}, (b)~\ch{RbMnF3}, (c)~\ch{KNiF3} and (d)~\ch{KMgF3}.
The temperature dependence of the $\varepsilon^{\textrm{lf}}_{0}$ for \ch{RbMnF3} has been adapted from Ref.~\cite{dubrovin2018unveiling}.
The color circles correspond to the experimental data.
The black and green lines are fits assuming the anharmonic and spontaneous magnetodielectric effects, respectively.
Values of the spontaneous magnetodielectric effect $\Delta\varepsilon^{\textrm{MD}}$ are given.
The paramagnetic and antiferromagnetic phases are shown in blue and red color filled backgrounds, respectively.
}
\end{figure*}

Figure~\ref{fig:spin_phonon}(d) shows the relation between the relative changes of bond angles ${\Delta\theta}/{\theta_{0}}$ and lengths ${\Delta{r}}/{r_{0}}$ estimated using the calculated ion displacements and the computed phonon frequency shifts ${\Delta\omega^{\textrm{SP}}}/{\omega^{\textrm{NM}}}$ in \ch{RbMnF3} and \ch{KNiF3}. 
There is a good agreement between the ${\Delta\theta}/{\theta_{0}}$ and ${\Delta\omega^{\textrm{SP}}}/{\omega^{\textrm{NM}}}$ for TO phonons in both crystals.
The smallest value of ${\Delta\theta}/{\theta_{0}}$ corresponds to the least ${\Delta\omega^{\textrm{SP}}}/{\omega^{\textrm{NM}}}$ for the 1TO phonon, while the highest frequency shift is observed for 2TO phonon, which has the most pronounced dynamic modulation of the superexchange bond angle as shown in Fig.~\ref{fig:spin_phonon}(d).
This result is in good agreement with the similar calculations in oxide perovskites~\cite{lee2011large}.
For the LO polar phonons the satisfactory agreement between ${\Delta\theta}/{\theta_{0}}$ and ${\Delta\omega^{\textrm{SP}}}/{\omega^{\textrm{NM}}}$ is observed only for 1LO phonon in both crystals, whereas the expected frequency shifts for calculated relative changes of bond angle ${\Delta\theta}/{\theta_{0}}$ for 2LO and 3LO phonons exceed the computed frequency shifts ${\Delta\omega^{\textrm{SP}}}/{\omega^{\textrm{NM}}}$ as can be seen in Fig.~\ref{fig:spin_phonon}(d).
% Apparently, this is due to ... \textcolor{red}{\textbf{Why?}}
Apparently, this is due to that the longitudinal ion displacements ${\Delta{r}}/{r_{0}}$ have a more significant effect which competes with bond angle changes ${\Delta\theta}/{\theta_{0}}$ on the spin-phonon coupling for LO phonons.
It is worth noting that the phonon-induced change bond lengths ${\Delta{r}}/{r_{0}}$ deviate markedly from the frequency shifts ${\Delta\omega^{\textrm{SP}}}/{\omega^{\textrm{NM}}}$ for TO phonons [see Fig.~\ref{fig:spin_phonon}(d)], which indicates that variation of the energy difference between orbitals $\Delta$ has a weaker effect on superexchange interaction $J$ than the change of the hopping integral $\tau$ in this case, what is consistent with published data~\cite{lee2011large,son2019unconventional}. 
Thereby, the frequency shifts $\Delta\omega^{\textrm{SP}}$ due to the spin-phonon coupling found in antiferromagnets \ch{RbMnF3} and \ch{KNiF3} are closely related to the dynamical modulation of the \ch{M}--\ch{F}--\ch{M} bond angle by TO phonons, while for LO phonons the noticeable competing effect of the bond length change is observed.

\subsection{Dielectric properties}
% \begin{figure*}[t]
% \centering
% \includegraphics[width=2\columnwidth]{fig_dielectric_20072021_compressed.pdf}
% %\includegraphics[width=2\columnwidth]{fig_dielectric_29062021.pdf}
% \caption{\label{fig:dielectric}
% Temperature dependences of (bottom frames) the dielectric strength $\Delta\varepsilon_{j}$ for $T_{1u}$ phonons $j=1-3$, (middle frames) the static dielectric permittivity $\varepsilon_{0}$, and (upper frames) the low-frequency dielectric permittivity $\varepsilon^{\textrm{lf}}_{0}$ at $f=100$\,kHz for the cubic fluoroperovskites (a)~\ch{KZnF3}, (b)~\ch{RbMnF3}, (c)~\ch{KNiF3} and (d)~\ch{KMgF3}.
% The temperature dependence of the $\varepsilon^{\textrm{lf}}_{0}$ for \ch{RbMnF3} has been adapted from Ref.~\cite{dubrovin2018unveiling}.
% The color circles correspond to the experimental data.
% The black and green lines are fits assuming the anharmonic and spontaneous magnetodielectric effects, respectively.
% Values of the spontaneous magnetodielectric effect $\Delta\varepsilon^{\textrm{MD}}$ are given.
% The paramagnetic and antiferromagnetic phases are shown in blue and red color filled backgrounds, respectively.
% }
% \end{figure*}

The experimental temperature dependences of the low-frequency dielectric permittivity $\varepsilon^{\textrm{lf}}_{0}(T)$ for the studied cubic fluoroperovskites are shown in the upper frames of Fig.~\ref{fig:dielectric}.
The data for \ch{RbMnF3} have been adapted from Ref.~\cite{dubrovin2018unveiling}.
The $\varepsilon^{\textrm{lf}}_{0}(T)$ mainly grows at cooling in \ch{KZnF3}, \ch{RbMnF3}, and \ch{KMgF3} crystals, while the more complex behavior was found in \ch{KNiF3}, in which a decrease of this quantity turns into an increase as the temperature is reduced.
Wherein, the relative changes of $\varepsilon^{\textrm{lf}}_{0}(T)$ are quite small and are about 1\% for \ch{RbMnF3}, \ch{KNiF3}, and \ch{KMgF3}, while for \ch{KZnF3} this value is about 5\% in the measured temperature range, which are quite typical for fluoroperovskites~\cite{dubrovin2018unveiling,dubrovin2019lattice,dubrovin2020spontaneous}.

The experimental frequencies of the polar phonons from Fig.~\ref{fig:phonon} allow us to obtain the dielectric strength $\Delta\varepsilon_{j}$ of a particular $j$th phonon from the expression~\cite{gervais1983long}
\begin{equation}
\label{eq:oscillator_strength_TOLO}
\Delta\varepsilon_{j}  =  \frac{\varepsilon_{\infty}}{{\omega^{2}_{j\textrm{TO}}}}\frac{\prod\limits_{k}{\omega^{2}_{k\textrm{LO}}}-{\omega^{2}_{j\textrm{TO}}}}{\prod\limits_{k\neq{}j}{\omega^{2}_{k\textrm{TO}}}-{\omega^{2}_{j\textrm{TO}}}},
\end{equation}
which corresponds to the contribution of this polar mode to the static dielectric permittivity $\varepsilon_{0} = \varepsilon_{\infty} + \sum_{j}\Delta\varepsilon_{j}$.
The calculated values of $\Delta\varepsilon$ and $\varepsilon_{0}$ from experiments at room temperature for the studied cubic fluoroperovskites are listed in Table~\ref{tab:phonon_parameters}.
Figure~\ref{fig:dielectric} shows the temperature dependences of the dielectric strengths $\Delta\varepsilon_{j}$ (bottom frames) and the static dielectric permittivity $\varepsilon_{0}(T)$ (middle frames).
The color circles correspond to the experimental data whereas the black and green lines are the fits including and neglecting the antiferromagnetic ordering below $T_{N}$, respectively.
There is a qualitative agreement between temperature dependences of the low-frequency $\varepsilon^{\textrm{lf}}_{0}$ and static $\varepsilon_{0}$ dielectric permittivities in the studied crystals, as seen in the upper and middle frames of Fig.~\ref{fig:dielectric}.
The temperature behaviors of $\Delta\varepsilon$ are very similar in the crystals under study and close to that previously observed in \ch{KCoF3} and \ch{RbCoF3}~\cite{dubrovin2019lattice}.
The $\Delta\varepsilon_{1}$ grows rapidly while the $\Delta\varepsilon_{2}$, on the contrary, decreases at cooling, whereas, the temperature changes of the $\Delta\varepsilon_{3}$ are quite insignificant in the all studied crystals.

The temperature dependences of the low-frequency dielectric permittivity $\varepsilon^{\textrm{lf}}_{0}$ exhibit kinks at $T_{N}$ in antiferromagnets \ch{RbMnF3} and \ch{KNiF3} due to the spontaneous magnetodielectric effect as shown in the upper frames in Figs.~\ref{fig:dielectric}(b) and~\ref{fig:dielectric}(c), respectively.
It should be noted that this effect was previously experimentally observed in some other magnetic fluoroperovskites with different crystal structures~\cite{dubrovin2018unveiling,dubrovin2019lattice,dubrovin2020incipient,dubrovin2020spontaneous}.
The frequency shifts $\Delta\omega^{\textrm{SP}}$ of the polar phonons caused by spin-phonon coupling lead to changes of the dielectric strengths $\Delta\varepsilon^{\textrm{MD}}$ as a result of the spontaneous magnetodielectric effect below $T_{N}$ as shown by the green lines in the bottom frames of Figs.~\ref{fig:dielectric}(b) and~\ref{fig:dielectric}(c).
The well-pronounced spontaneous magnetodielectric effect $\Delta\varepsilon^{\textrm{MD}}$ is observed for $\Delta\varepsilon_{2}$ and $\Delta\varepsilon_{3}$, and has negative sign in both crystals.
It is worth noting that the absolute values of $\Delta\varepsilon^{\textrm{MD}}$ are fairly close in these crystals even though the spin-phonon coupling is more pronounced in \ch{KNiF3} than in \ch{RbMnF3}.
This is due to that the value of $\Delta\varepsilon^{\textrm{MD}}$ of the polar phonon is determined by the relative changes of $\omega_{\textrm{TO}}$ and $\omega_{\textrm{LO}}$ caused by frequency shifts $\Delta\omega^{\textrm{SP}}$ not only this, but also others polar phonons according to Eq.~\eqref{eq:oscillator_strength_TOLO}.

A good qualitative agreement is observed in changes of temperature behavior of the low-frequency $\varepsilon^{\textrm{lf}}_{0}$ and the static $\varepsilon_{0}$ dielectric permittivities due to the spontaneous magnetodielectric effect below $T_{N}$ in both crystals, as it can be seen in the upper and middle frames in Figs.~\ref{fig:dielectric}(b) and~\ref{fig:dielectric}(c).
It should be noted that the signs and relative values of the spontaneous magnetodielectric effect observed in the experiments are in satisfactory agreement with $\Delta\varepsilon^{\textrm{MD}} = \Delta\varepsilon^{\textrm{AFM}} - \Delta\varepsilon^{\textrm{NM}}$ obtained by using Eq.~\eqref{eq:oscillator_strength_TOLO} with $\omega^{\textrm{AFM}}$ and $\omega^{\textrm{NM}}$ phonon frequencies from our DFT simulations for both antiferromagnetic crystals, as presented in Table~\ref{tab:dft_SP}.
Remarkably, we have shown that the microscopic lattice dynamics features in the cubic fluoroperovskites, such as the softening polar phonons and frequency shifts due to the spin-phonon coupling, manifest themselves in macroscopic effects, which can be effectively studied using the low-frequency dielectric spectroscopy. 

% \subsection{Effective plasma frequency}
% \begin{figure*}
% \centering
% \includegraphics[width=2\columnwidth]{fig_plasma_06042020.pdf}
% \caption{\label{fig:plasma}
% Temperature dependences of (bottom frames) the effective ionic plasma frequencies $\Omega_{j\textrm{P}}$ for $T_{1u}$ phonons $j$=1--3, and (upper frames) total frequency $\Omega_{\Sigma\textrm{P}}$ of cubic fluoroperovskites (a)~\ch{KZnF3}, (b)~\ch{RbMnF3}, (c)~\ch{KNiF3} and (d)~\ch{KMgF3}.
% The color circles correspond to the experimental data.
% The black and green lines are fits assuming the anharmonic effects and magnetic ordering, respectively.
% The paramagnetic and antiferromagnetic phases are shown in blue and red color filled backgrounds, respectively.
% }
% \end{figure*}

% The dielectric strength $\Delta\varepsilon_{j}$ can by related to an effective ionic plasma frequency $\Omega^{2}_{j\textrm{P}} = \Delta\varepsilon_{j}\omega^{2}_{j\textrm{TO}}$ of $j$th phonon~\cite{scott1971raman}.
% The effective ionic plasma frequency is related to the \mbox{LO-TO} splitting and effective ionic charges as described in detail in Refs.~\cite{scott1971raman,gervais1983long,dubrovin2019lattice}.

\section{Conclusions}
In summary, we have systematically studied the lattice dynamics of the cubic fluoroperovskites by far-infrared reflectivity technique and supported by the first-principles calculations.
We have experimentally demonstrated that the polar phonons at the $\Gamma$ point of the Brillouin zone, which are stable for all studied crystals, are softening by several cm$^{-1}$ at cooling, indicating the proximity of studied crystals to incipient ferroelectrics.
We revealed that the harmonic $k_{0}$ and anharmonic $k_{\textrm{ah}}$ force constants associated with these softening polar phonons are mutually coupled and correlate with the tolerance factor $t$ of the cubic fluoroperovskites. 
Furthermore, we observed that as the lower is the $t$, the smaller is the value of $k_{0}$ and the most significant is the temperature change of $k_{\textrm{ah}}$.
Thus, the disclosed trend leads to that the lower is the value of the tolerance factor $t$ of the studied cubic fluoroperovskite, the closer its anharmonic properties become to incipient ferroelectrics.
However, according to our simulations, epitaxial strain does not lead to the ferroelectric instability and, hence, cubic fluoroperovskites are not incipient ferroelectrics.

Based on our first-principles calculations, we disclosed the physical origin of the tolerance factor $t$ in the cubic fluoroperovskites.
According to our lattice dynamics simulations, the computed harmonic force constants $k$ of the lowest-frequency phonons tend to decrease with a reduction of $t$ not only at the $\Gamma$ point, but at all high-symmetry points of the Brillouin zone.
Thus, the revealed trends point to the incipient lattice instability in the cubic fluoroperovskites which is realized only in the crystals with small values of $t$ and drives to phase transitions from the cubic to the tetragonal and orthorhombic structures.
However, these transitions lead to nonpolar structures but not to polar ones, since the phonon condensation that caused it occurs rather at the $M$ and $R$ points than at the $\Gamma$ point of the Brillouin zone.
The disclosed correlation with the tolerance factor $t$ indicates the geometric origin of the observed incipient lattice instability caused by the steric effect due to the unit-cell volume filling by ions rather than forming a strong covalent bond as in oxide perovskites.

We found that the frequency shifts due to spin-phonon coupling observed in the antiferromagnetic cubic fluoroperovskites can be understood in terms of the dynamical modulation of the \ch{M}--\ch{F}--\ch{M} bond angle induced by the relevant TO polar phonons, while the competing effect from bond length modulation is noticeable for the LO polar phonons.
Finally, we have shown that low-frequency dielectric spectroscopy fairly reflects the observed lattice dynamics features such as the softening of polar phonons and frequency shifts due to the spin-phonon coupling in the studied insulating crystals. 

We believe that our results will stimulate further experimental and theoretical studies of the unusual lattice dynamics of highly anharmonic inorganic metal halide perovskites.
These efforts will be relevant to further optimization of their physical properties for the rational design of multifunctional devices. 
Moreover, we can envisage that the cubic fluoroperovskites with the different influence of the antiferromagnetic ordering on the TO and LO polar phonons are close to becoming model materials for the rapidly developing field of THz magnetophononics~\cite{stupakiewicz2020ultrafast,afanasiev2021ultrafast}.

\section*{Acknowledgments}
The single crystals provided by J.-Y.\,Gesland, P.\,P.\,Syrnikov, and S.\,V.\,Petrov were used in experiments.
We are grateful to O.\,A.\,Alekseeva and M.\,P.\,Scheglov for the help with x-ray orientation and preparation of samples, and D.\,A.\,Andronikova, A.\,S.\,Sheremet, R.\,G.\,Burkovsky, and A.\,K.\,Tagantsev for fruitful scientific discussions.
% We are grateful to J.-Y.\,Gesland, P.\,P.\,Syrnikov and S.\,V.\,Petrov for providing single crystals, O.\,A.\,Alekseeva and M.\,P.\,Scheglov for the help with X-ray orientation and preparation of samples, and D.\,A.\,Andronikova, A.\,S.\,Sheremet, R.\,G.\,Burkovsky and A.\,K.\,Tagantsev for fruitful scientific discussions.
R.\,M.\,D., N.\,V.\,S., and R.\,V.\,P. acknowledge the support by Russian Foundation for Basic Research according to the project No.\,19-02-00457.
N.\,N.\,N. and K.\,N.\,B. acknowledge the Ministry of Science and Higher Education of Russia under the grant No.\,0039-2019-0004.
Calculations presented in this work were carried out using the GridUIS-2 experimental testbed, being developed under the Universidad Industrial de Santander (SC3-UIS) High Performance and Scientific Computing Centre, development action with support from UIS Vicerrector\'ia de Investigaci\'on y Extensi\'on (VIE-UIS) and several UIS research groups as well as other funding resources.
Additionally, we acknowledge the XSEDE facilities' support, a project from the National Science Foundation under grant number ACI-1053575.
The authors also acknowledge the Texas Advanced Computer Center (with the Stampede2 and Bridges-2 supercomputers).
We also acknowledge the use of the SuperComputing System (Thorny Flat) at WVU, which is funded in part by the National Science Foundation (NSF) Major Research Instrumentation Program (MRI) Award $\#$1726534.
A.\,C.\,G.\,C. acknowledges the grant No. 2677: ``Quiralidad y Ordenamiento Magnético en Sistemas Cristalinos: Estudio Teórico desde Primeros Principios'' supported by the VIE – UIS.
A.\,H.\,R. acknowledges the support of DMREF-NSF 1434897 and NSF OAC-1740111 projects.

\bibliography{bibliography}

\begin{table*}[h]
    \caption{\label{tab:phonon_parameters} Experimental lattice parameters $a$ (\AA), frequencies $\omega_{j}$ (cm$^{-1}$), and dielectric strengths $\Delta\varepsilon_{j}$ of the IR-active phonons, static $\varepsilon_{0}$ and high frequency $\varepsilon_{\infty}$ dielectric permittivities at room temperature in the cubic fluoroperovskites with different tolerance factors $t$.}
    \begin{ruledtabular}
            \begin{tabular}{cccccccccccccccccccc}
             \ch{AMF3} & $t$ & $a$ & $\omega_{1\textrm{TO}}$ & $\gamma_{1\textrm{TO}}$ & $\omega_{1\textrm{LO}}$ & $\gamma_{1\textrm{LO}}$ & $\omega_{2\textrm{TO}}$ & $\gamma_{2\textrm{TO}}$ & $\omega_{2\textrm{LO}}$ & $\gamma_{2\textrm{LO}}$ & $\omega_{3\textrm{TO}}$ & $\gamma_{3\textrm{TO}}$ & $\omega_{3\textrm{LO}}$ & $\gamma_{3\textrm{LO}}$ & $\varepsilon_{0}$ & $\Delta\varepsilon_{1}$ & $\Delta\varepsilon_{2}$ & $\Delta\varepsilon_{3}$ & $\varepsilon_{\infty}$\\
             \hline
             \ch{KZnF3}  & 0.95 & 4.055\footnote{Refs.~\cite{knight2017low,knox1961perovskite}.} & 139.9 & 5.0 & 150.1 & 5.3 & 195.6 & 8.5  & 301.3 & 7.1 & 408.1 & 37.3 & 492.0 & 30.8 & 8.6 & 1.89 & 3.98 & 0.57 & 2.17\\
             \ch{RbMnF3} & 0.96 & 4.2396\footnote{Ref.~\cite{windsor1966spin}.} & 112.4 & 6.1 & 123.6 & 6.7 & 194.4 & 14.1 & 269.8 & 7.4 & 379.3 & 23.5 & 458.1 & 22.7 & 7.1 & 1.57 & 2.81 & 0.63 & 2.09\\
             \ch{KNiF3}  & 0.96 & 4.014\footnote{Refs.~\cite{okazaki1961crystal,knox1961perovskite}.}  & 149.6 & 7.6 & 165.3 & 7.5 & 246.0 & 13.1 & 305.7 & 9.5 & 447.5 & 25.6 & 527.7 & 38.0 & 5.61 & 1.27 & 1.58 & 0.62 & 2.14\\
             \ch{KMgF3}  & 0.97 & 4.006\footnote{Ref.~\cite{vaitheeswaran2007high}.}  & 166.9 & 8.9 & 195.7 & 4.4 & 299.7 & 7.8  & 358.8 & 5.8 & 456.7 & 17.1 & 555.4 & 29.8 & 5.39 & 1.75 & 1.23 & 0.56 & 1.85\\
        \end{tabular}
    \end{ruledtabular}
\end{table*}

\begin{table*}[h]
    \caption{\label{tab:dft_phonon_parameters} Calculated lattice parameters $a$, frequencies $\omega_{j}$ (cm$^{-1}$), and  dielectric strengths $\Delta\varepsilon_{j}$ of the IR-active phonons, static $\varepsilon_{0}$ and high frequency $\varepsilon_{\infty}$ dielectric permittivities in the cubic fluoroperovskites with different tolerance factors $t$.
    For magnetic crystals, the calculations were performed at the antiferromagnetic spin ordering.}
    \begin{ruledtabular}
            \begin{tabular}{cccccccccccccc}
             \ch{AMF3} & $t$ & $a$ (\AA) & $\omega_{1\textrm{TO}}$ & $\omega_{1\textrm{LO}}$ & $\omega_{2\textrm{TO}}$ & $\omega_{2\textrm{LO}}$ & $\omega_{3\textrm{TO}}$ & $\omega_{3\textrm{LO}}$ & $\varepsilon_{0}$ & $\Delta\varepsilon_{1}$ & $\Delta\varepsilon_{2}$ & $\Delta\varepsilon_{3}$ & $\varepsilon_{\infty}$\\
             \hline
             \ch{RbCaF3} & 0.88 & 4.436 & 63.1  & 104.8 & 180.1 & 226.1 & 397.0 & 484.9 & 14.31 & 9.68 & 1.55 & 0.88 & 2.20\\
             \ch{KMnF3}  & 0.91 & 4.193 & 94.5  & 132.0 & 181.7 & 244.2 & 425.0 & 501.6 & 11.36 & 6.56 & 1.79 & 0.71 & 2.31\\
             \ch{KCoF3}  & 0.94 & 4.039 & 134.3 & 149.7 & 213.1 & 273.0 & 432.0 & 517.3 & 7.05 & 1.79 & 2.04 & 0.81 & 2.41\\
             \ch{KZnF3}  & 0.95 & 4.067 & 128.7 & 139.4 & 180.9 & 273.7 & 414.7 & 484.8 &  8.82 & 2.11 & 3.70 & 0.60 & 2.40\\
             \ch{RbMnF3} & 0.96 & 4.237 & 105.4 & 117.8 & 187.8 & 253.7 & 390.2 & 465.3 &  7.89 & 1.94 & 2.76 & 0.76 & 2.44\\
             \ch{KNiF3}  & 0.96 & 4.002 & 144.8 & 160.5 & 239.0 & 287.7 & 438.8 & 514.3 &  6.32 & 1.44 & 1.54 & 0.76 & 2.58\\
             \ch{KMgF3}  & 0.97 & 3.999 & 156.9 & 188.6 & 283.1 & 341.4 & 462.5 & 553.2 &  6.35 & 2.32 & 1.30 & 0.62 & 2.11\\
             \ch{RbCoF3} & 1.00 & 4.095 & 124.9 & 129.6 & 224.9 & 282.0 & 381.7 & 466.8 &  6.48 & 0.56 & 2.49 & 0.87 & 2.56\\
        \end{tabular}
    \end{ruledtabular}
\end{table*}

% % This Table is not used in the main text.
% \begin{table}[h]
%     \caption{\label{tab:dft_force_constants} Calculated force constants $k_{j}$ (eV/\AA$^2$) of IR active phonons at the Brillouin zone center and lattice parameters $a$ in cubic fluoroperovskites with different tolerance factors $t$.}
%     \begin{ruledtabular}
%             \begin{tabular}{ccccccccc}
%              \ch{AMF3} & $t$ & $a$ (\AA) & $k_{1\textrm{TO}}$ & $k_{1\textrm{LO}}$ & $k_{2\textrm{TO}}$ & $k_{2\textrm{LO}}$ & $k_{3\textrm{TO}}$ & $k_{3\textrm{LO}}$ \\
%              \hline
%              \ch{RbCaF3} & 0.88 & 4.436 &  4.5 & 15.6 & 21.7 & 25.7 &  94.9 & 152.0\\
%              \ch{KMnF3}  & 0.91 & 4.193 &  6.9 & 16.9 & 24.4 & 31.8 & 111.4 & 170.8\\
%              \ch{KCoF3}  & 0.94 & 4.039 & 15.7 & 22.4 & 30.7 & 39.8 & 116.3 & 183.1\\
%              \ch{KZnF3}  & 0.95 & 4.067 & 13.8 & 20.3 & 26.2 & 40.7 & 100.8 & 161.6\\
%              \ch{RbMnF3} & 0.96 & 4.237 & 15.2 & 25.6 & 24.7 & 35.2 &  93.5 & 150.7\\
%              \ch{KNiF3}  & 0.96 & 4.002 & 18.4 & 21.4 & 36.8 & 43.8 & 124.3 & 184.1\\
%              \ch{KMgF3}  & 0.97 & 3.999 & 19.1 & 27.8 & 40.5 & 55.6 & 106.4 & 159.9\\
%              \ch{RbCoF3} & 1.00 & 4.095 & 24.0 & 25.8 & 33.0 & 45.3 &  89.5 & 154.4\\
%         \end{tabular}
%     \end{ruledtabular}
% \end{table}

\begin{table}[h]
    \caption{\label{tab:dft_born_charges} Calculated Born effective charges in units of the elementary charge $e$ in the cubic fluoroperovskites with different tolerance factors $t$.
    $\textrm{F}_{\perp}$ and $\textrm{F}_{\parallel}$ indicate the Born effective charge when it is displaced perpendicular and parallel to the \ch{M}--\ch{F}--\ch{M} bond.} 
    \begin{ruledtabular}
            \begin{tabular}{cccccc}
             \ch{AMF3} & $t$ & $Z_{\textrm{A}}$ & $Z_{\textrm{M}}$ & $Z_{\textrm{F}\perp}$ & $Z_{\textrm{F}\parallel}$\\
             \hline
             Nominal     &      & 1    & 2    & -1    & -1\\
             \ch{RbCaF3} & 0.88 & 1.22 & 2.37 & -0.97 & -1.67\\
             \ch{KMnF3}  & 0.91 & 1.20 & 2.28 & -0.88 & -1.71\\
             \ch{KCoF3}  & 0.94 & 1.20 & 2.21 & -0.83 & -1.86\\
             \ch{KZnF3}  & 0.95 & 1.20 & 2.30 & -0.90 & -1.70\\
             \ch{RbMnF3} & 0.96 & 1.25 & 2.33 & -0.94 & -1.70\\
             \ch{KNiF3}  & 0.96 & 1.21 & 2.15 & -0.80 & -1.75\\
             \ch{KMgF3}  & 0.97 & 1.19 & 2.00 & -0.96 & -1.27\\
             \ch{RbCoF3} & 1.00 & 1.27 & 2.28 & -0.89 & -1.85\\
        \end{tabular}
    \end{ruledtabular}
\end{table}

% \begin{table}[h]
%     \caption{\label{tab:dft_eigendisplacements} Calculated eigendisplacements of the low-frequency $T_{1u}$ mode $\nu$ in cubic fluoroperovskites with different tolerance factors $t$.}
%     \begin{ruledtabular}
%             \begin{tabular}{cccccc}
%              \ch{AMF3} & $t$ & \ch{A} & \ch{M} & \ch{F}$_{\perp}$ & \ch{F}$_{\parallel}$\\
%              \hline
%              \ch{RbCaF3} & 0.88 & 0.076428 & -0.046159 & -0.097516 & -0.053334\\
%              \ch{KMnF3}  & 0.91 & 0.134272 & -0.029046 & -0.075981 & -0.039332\\
%              \ch{KCoF3}  & 0.94 & 0.138834 & -0.050146 & -0.037559 & -0.048446\\
%              \ch{KZnF3}  & 0.95 & 0.139284 & -0.044716 & -0.043609 & -0.045014\\
%              \ch{RbMnF3} & 0.96 & 0.081428 & -0.061823 & -0.062186 & -0.063407\\
%              \ch{KNiF3}  & 0.96 & 0.137637 & -0.052806 & -0.034893 & -0.050993\\
%              \ch{KMgF3}  & 0.97 & 0.130925 & -0.049887 & -0.072064 & -0.061555\\
%              \ch{RbCoF3} & 1.00 & 0.080942 & -0.073719 & -0.032151 & -0.065094\\
%         \end{tabular}
%     \end{ruledtabular}
% \end{table}

\begin{table*}[h]
    \caption{\label{tab:dft_phonons_BZ} Calculated frequencies $\omega$ (cm$^{-1}$) and force constants $k$ (eV/\AA$^2$) of the low-lying $T_{1u}$, $X_5$, $M_{2}$, and $R_{15'}$ phonons at the $\Gamma$, $X$, $M$, and $R$ points of the Brillouin zone, respectively, in the cubic fluoroperovskites with different tolerance factors $t$.}
    \begin{ruledtabular}
            \begin{tabular}{ccccccccccccccccc}
                         &      &          && \multicolumn{2}{c}{$T_{1u}$} && \multicolumn{2}{c}{$X_{5}$} && \multicolumn{2}{c}{$M_{2}$} && \multicolumn{2}{c}{$R_{15'}$}\\ \cmidrule{5-6} \cmidrule{8-9} \cmidrule{11-12} \cmidrule{14-15}
             \ch{AMF3}   & $t$  & $a$ (\AA) && $\omega$ & $k$ && $\omega$ & $k$ && $\omega$ & $k$ && $\omega$ & $k$\\\hline
             \ch{RbCaF3} & 0.88 & 4.436 &&  63.1 &  4.5 && 51.5  & 2.46 && $i$43.6 & -0.86 && $i$49.1 & -1.09\\
             \ch{KMnF3}  & 0.91 & 4.193 &&  94.5 &  6.9 && 74.2  & 3.27 && $i$47.4 & -1.02 && $i$51.2 & -1.19\\
             \ch{KCoF3}  & 0.94 & 4.039 && 134.3 & 15.7 && 92.7  & 6.04 && 75.5  & 2.59 && 72.4  & 2.38\\
             \ch{KZnF3}  & 0.95 & 4.067 && 128.7 & 13.8 && 89.7  & 5.72 && 64.2  & 1.87 && 60.5  & 1.66\\
             \ch{RbMnF3} & 0.96 & 4.237 && 105.4 & 15.2 && 76.3  & 5.62 && 76.7  & 3.58 && 84.3  & 3.23\\
             \ch{KNiF3}  & 0.96 & 4.002 && 144.8 & 18.4 && 111.3 & 8.57 && 97.3  & 4.30 && 94.3  & 4.04\\
             \ch{KMgF3}  & 0.97 & 3.999 && 156.9 & 19.1 && 125.7 & 8.17 && 101.3 & 4.67 && 98.1  & 4.37\\
             \ch{RbCoF3} & 1.00 & 4.095 && 124.9 & 24.0 && 95.9  & 8.26 && 96.7  & 8.89 && 111.0 & 8.4\\
        \end{tabular}
    \end{ruledtabular}
\end{table*}

% \begin{table}[h]
%     \caption{\label{tab:dft_SP} Frequencies $\omega$ (cm$^{-1}$) and force constants $k$ (eV/\AA$^2$) of the low-lying optical at the $\Gamma$ point and acoustic phonons at the $X$, $R$ and $M$ points of the Brillouin zone in the cubic fluoroperovskites \ch{AMF3}.}
%     \begin{ruledtabular}
%             \begin{tabular}{ccccccccccc}
%                  && \multicolumn{4}{c}{\ch{RbMnF3}} && \multicolumn{4}{c}{\ch{KNiF3}}\\ \cmidrule{3-6} \cmidrule{8-11}
%              Mode && $\omega^{\textrm{AFM}}$ & $\omega^{\textrm{FM}}$ & $\omega^{\textrm{SP}}$ & $\Delta\varepsilon^{\textrm{MD}}$ && $\omega^{\textrm{AFM}}$ & $\omega^{\textrm{FM}}$ & $\omega^{\textrm{SP}}$ & $\Delta\varepsilon^{\textrm{MD}}$\\
%              1TO && 105.4 & 105.4 & 0.0  & X.XX && 144.8 & 144.6 &  0.1  & X.XX\\
%              1LO && 117.8 & 117.4 & 0.2  & X.XX && 160.5 & 156.4 &  2.05 & X.XX\\
%              2TO && 187.8 & 184.8 & 1.5  & X.XX && 239.0 & 218.3 & 10.35 & X.XX\\
%              2LO && 253.7 & 252.6 & 0.55 & X.XX && 287.7 & 280.5 &  3.6  & X.XX\\
%              3TO && 390.2 & 389.3 & 0.45 & X.XX && 438.8 & 437.2 &  0.8  & X.XX\\
%              3LO && 465.3 & 465.3 & 0.0  & X.XX && 514.3 & 519.0 & -2.35 & X.XX\\
%         \end{tabular}
%     \end{ruledtabular}
% \end{table}

\begin{table}[h]
    \caption{\label{tab:dft_SP} Computed polar phonon frequencies $\omega$ (cm$^{-1}$) in G-AFM, FM and NM states, spin-phonon coupling $\Delta\omega^{\textrm{SP}}$ (cm$^{-1}$) and spontaneous magnetodielectric effect $\Delta\varepsilon^{\textrm{MD}}_{0}$ coefficients for the cubic fluoroperovskites \ch{RbMnF3} and \ch{KNiF3}.}
    \begin{ruledtabular}
            \begin{tabular}{ccccccc}
             Mode && $\omega^{\textrm{AFM}}$ & $\omega^{\textrm{FM}}$ & $\omega^{\textrm{NM}}$ & $\Delta\omega^{\textrm{SP}}$ & $\Delta\varepsilon^{\textrm{MD}}$ \\ \cmidrule{3-7}
                 && \multicolumn{5}{c}{\ch{RbMnF3}} \\
             1TO && 105.4 & 105.4 & 105.4 & 0.0  & \multirow{2}{*}{-0.007} \\ 
             1LO && 117.8 & 117.4 & 117.6 & 0.2  &  \\
             2TO && 187.8 & 184.8 & 186.3 & 1.5  & \multirow{2}{*}{-0.073} \\
             2LO && 253.7 & 252.6 & 253.2 & 0.55 &  \\
             3TO && 390.2 & 389.3 & 389.8 & 0.45 & \multirow{2}{*}{-0.004} \\
             3LO && 465.3 & 465.3 & 465.3 & 0.0  &  \\
                 &&       &       &       & \multicolumn{2}{c}{$\Delta\varepsilon^{\textrm{MD}}_{0}=-0.084$} \\ \hline
                 && \multicolumn{5}{c}{\ch{KNiF3}} \\
             1TO && 144.8 & 144.6 & 144.7 &  0.1  & \multirow{2}{*}{0.021} \\
             1LO && 160.5 & 156.4 & 158.5 &  2.05 & \\
             2TO && 239.0 & 218.3 & 228.7 & 10.35 & \multirow{2}{*}{-0.332} \\
             2LO && 287.7 & 280.5 & 284.1 &  3.6  & \\
             3TO && 438.8 & 437.2 & 438.0 &  0.8  & \multirow{2}{*}{-0.025} \\
             3LO && 514.3 & 519.0 & 516.7 & -2.35 & \\
                 &&       &       &       & \multicolumn{2}{c}{$\Delta\varepsilon^{\textrm{MD}}_{0}=-0.336$} \\
            % \hline
            %      && \multicolumn{5}{c}{\ch{KCoF3}} \\  
            %  1TO && 134.3 & 133.9 & 134.1 &  0.2  & \multirow{2}{*}{-0.071} \\
            %  1LO && 149.7 & 153.8 & 151.8 &  2.05 & \\ 
            %  2TO && 213.1 & 234.9 & 224.0 & -10.9 & \multirow{2}{*}{0.269} \\
            %  2LO && 273.0 & 284.5 & 278.8 & -5.75 & \\ 
            %  3TO && 432.0 & 437.3 & 434.7 & -2.65 & \multirow{2}{*}{-0.005} \\ 
            %  3LO && 517.3 & 523.7 & 520.5 & -3.2  & \\
            %      &&       &       &       & \multicolumn{2}{c}{$\Delta\varepsilon^{\textrm{MD}}_{0}$ = 0.193} \\ \hline
            %      && \multicolumn{5}{c}{\ch{RbCoF3}} \\
            %  1TO && 124.9 & 126.4 & 125.7 & -0.75 & \multirow{2}{*}{-0.1137} \\
            %  1LO && 129.6 & 133.9 & 131.8 & -2.15 & \\
            %  2TO && 224.9 & 244.6 & 234.8 & -9.85 & \multirow{2}{*}{0.3056} \\
            %  2LO && 282.0 & 292.2 & 287.1 & -5.1  & \\
            %  3TO && 381.7 & 387.5 & 384.6 & -2.9  & \multirow{2}{*}{-0.0125} \\
            %  3LO && 466.8 & 474.3 & 470.6 & -3.75 & \\
            %      &&       &       &       & \multicolumn{2}{c}{$\Delta\varepsilon^{\textrm{MD}}_{0}$ = 0.180} \\
        \end{tabular}
    \end{ruledtabular}
\end{table}

\end{document}